\documentclass[aps,prc,twocolumn,times,graphicx,tighten,showpacs]{revtex4}
\usepackage[dvips]{graphicx}
\usepackage[dvips]{color}

\newcommand{\bea}{\begin{eqnarray}}
\newcommand{\eea}{\end{eqnarray}}
\newcommand{\be}{\begin{equation}}
\newcommand{\ee}{\end{equation}}
\newcommand{\np}{{\bf p}}
\newcommand{\hp}{\widehat{\bf p}}
\newcommand{\ta}{\tilde{a}}
\newcommand{\tb}{\tilde{b}}
\newcommand{\tv}{\tilde{v}}
\newcommand{\nh}{{\bf h}}
\newcommand{\nk}{{\bf k}}
\newcommand{\nl}{{\bf l}}
\newcommand{\nx}{{\bf x}}
\newcommand{\nq}{{\bf q}}
\newcommand{\ns}{{\bf s}}

\begin{document}

\title{
Relativistic effects in two-particle emission 
for electron and neutrino reactions
}

\author{
I. Ruiz Simo$^a$,
C. Albertus$^a$,
J.E. Amaro$^a$,
M.B. Barbaro$^b$,
J.A. Caballero$^c$,
T.W. Donnelly$^d$,
}

\affiliation{$^a$Departamento de F\'{\i}sica At\'omica, Molecular y Nuclear,
and Instituto de F\'{\i}sica Te\'orica y Computacional Carlos I,
Universidad de Granada, Granada 18071, Spain}

\affiliation{$^b$Dipartimento di Fisica, Universit\`a di Torino and
  INFN, Sezione di Torino, Via P. Giuria 1, 10125 Torino, Italy}

\affiliation{$^c$Departamento de F\'{\i}sica At\'omica, Molecular y Nuclear,
Universidad de Sevilla, Apdo.1065, 41080 Sevilla, Spain}

\affiliation{$^d$Center for Theoretical Physics, Laboratory for Nuclear
  Science and Department of Physics, Massachusetts Institute of Technology,
  Cambridge, MA 02139, USA}

\date{\today}


\begin{abstract}
Two-particle two-hole contributions to electroweak response functions
are computed in a fully relativistic Fermi gas, assuming that the
electroweak current matrix elements are independent of the kinematics.
We analyze the genuine kinematical and relativistic effects before
including a realistic meson-exchange current (MEC) operator.  This
allows one to study the mathematical properties of the non-trivial
seven-dimensional integrals appearing in the calculation and to design
an optimal numerical procedure to reduce the computation time. This is
required for practical applications to CC neutrino scattering
experiments, where an additional integral over the neutrino flux is
performed.  Finally we examine the viability  of this model to compute the 
electroweak 2p-2h response functions.

\end{abstract}


\maketitle

\section{Introduction}

The understanding of intermediate-energy (0.5--10 GeV)
neutrino-nucleus scattering cross sections is an important ingredient
to atmospheric and accelerator-based neutrino oscillation
experiments~\cite{Gal11,For12,Mor12,Alv14}.  The analysis of
  these experiments requires having good control of nuclear
  effects. The simple
description based on a relativistic Fermi gas (RFG) model does not
accurately describe the recent measurements of quasielastic neutrino
and antineutrino scattering~\cite{Agu10,Agu13,Fio13,Abe13}.
Mechanisms such as nuclear correlations, final-state interactions and
Meson-exchange currents (MEC) may have an impact on the inclusive
neutrino charged current (CC) cross section. In particular, explicit
calculations support the theoretical evidence~\cite{Mar09,Nie11,Ama11}
for a significant contribution from multi-nucleon knock-out to the CC
cross sections $(\nu_{\mu},\mu^-)$ and $({\overline\nu}_{\mu},\mu^+)$
around and above the quasielastic (QE) peak region, defined by
$\omega=\sqrt{q^2+m_N^2}-m_N$, where $\omega$ is the energy transfer and
$q$ is the 3-momentum transfer.  Recent {\em ab initio}
calculations~\cite{Lov14} of sum rules of weak neutral-current
response functions of $^{12}$C have also stressed the importance of
MEC in neutrino quasielastic scattering. The size of MEC effects is
larger than that found in inclusive CC neutrino scattering from the
deuteron~\cite{Shen12}.

The three existing microscopic models that have provided predictions
of multi-nucleon knockout effects in quasielastic neutrino and
antineutrino cross sections from $^{12}$C for the experimental
kinematical settings are those by
Martini~\cite{Mar10,Mar11,Mar12,Mar13,Mar13b,Mar14},
Nieves~\cite{Nie11,Nie12,Nie13,Gra13}, and the Super-Scaling Analysis
(SuSA) model of~\cite{Ama11,Ama12,Ama13}.  

These three models are based on the Fermi gas but each one contains
different ingredients and approximations to face the problem. The
Martini model is based on the non-relativistic model of \cite{Alb84}
although attempts to improve it using relativistic kinematics have
been made. The model includes MEC and pionic correlation diagrams
modified to account for the effective nuclear interaction.  The
interference between direct and exchange diagrams is neglected, in
order to reduce the 7D integral over the phase space to a 2D
integration.  The Nieves model is similar to Martini's, but most of it
is fully relativistic. In this model the momentum of the initial
nucleon in the generic $WNN\pi$ vertex is fixed to an average
value. Under this approximation the Lindhard function can be
factorized inside the integral, leaving only a 4D integration over the
momentum of one of the exchanged pions.  The direct-exchange
interference is neglected as well.  The SuSA model includes all the
interference terms at the cost of performing a 7D integration, without
any approximation, but the axial part of the MEC is not yet included.
It is obvious that these three models should differ numerically
because they are different. But a quantitative evaluation of their
differences has not been done. Furthermore, the accuracy of the
approximations used in these models only can be determined by
comparison with an exact calculation for some kinematics.

Alternatively, phenomenological approaches have been proposed where 2p-2h
effects, estimated by a pure two-nucleon phase-space model, are
fitted to the experimental cross section~\cite{Lal12,Lal12b}, while
the nucleon ejection model of~\cite{Sob12} provides a phase-space based 
algorithm to generate 2p-2h states in a Monte Carlo implementation. 

The present paper is a first step towards an extension of the
relativistic 2p-2h model of~\cite{DePace03} to the weak sector.  We
undertake this project with the final goal of including a consistent
set of weak MEC in the SuSA approach to CC neutrino
reactions~\cite{Ama11,Ama12}.  The model of~\cite{DePace03} fully
described the contribution of 2p-2h states to the transverse response
function in electron scattering.  Based on the RFG, the model included
all 2p-2h diagrams containing two pionic lines (except for nucleon
correlations that were included in \cite{Ama10a}), taking into account
the quantum interferences between direct and exchange two-body matrix
elements.  Previous calculations of two-particle emission with MEC in
$(e,e')$ involved non-relativistic
models~\cite{Donnelly:1978xa,Van80,Alb84,Alb91,Ama93,Ama94,Gil97}.
The first attempts for a relativistic description were made by
Dekker~\cite{Dek91,Dek92,Dek94}, followed by the model of De Pace {\it
  et al.}~\cite{DePace03,DePace04}.  The extension of this model to
the weak sector requires the inclusion of the axial terms of
MEC. Quasielastic neutrino scattering requires one to perform an
integral over the neutrino flux. This would considerably increase the
computing-time of the nuclear response function of~\cite{DePace03}
involving 7D integrals of thousands of terms, although improvements
were made in \cite{Ama10a} to perform the spin traces numerically.
Thus in this work we address the problem from a different perspective,
focusing first on a careful study of the 7D integral over the 2p-2h
phase space as a function of the momentum and energy transfers.  Our
goal is to provide a comprehensive description of the angular
distribution, showing that there is a divergence in the integrand for
some kinematics, and identifying mathematically the allowed
integration intervals. At the same time we derive a procedure to
integrate the angular distribution around the divergence analytically.
This procedure allows us to reduce the CPU time considerably.  This
program is followed first in a pure phase-space domain, without yet
including the two-body current. We also sketch the future perspectives
opened by this general formalism applied to the calculation of 2p-2h
contributions to electroweak responses.  In a forthcoming paper, we
will provide a full model of weak MEC to compute the complete set of
CC neutrino scattering response functions.

The paper is organized as follows. In Sect. II we define the
relativistic 2p-2h response and phase space functions.  In Sect. III
we review the non-relativistic description of the 2p-2h integrals,
semi-analytical expressions that will be used as a check of the
relativistic calculations, and some interesting properties of the
phase-space integral, such as scaling and asymptotic expansion. In
Sect. IV we address the relativistic phase-space function and
asymptotic expansion, and show that some numerical problems arise from
a straightforward calculation for high $q$. In Sect. V we describe the
2p-2h angular distribution in the {\sl ``frozen nucleon''}
approximation and show that this distribution has a divergence for
some angles. The divergence is related to the two solutions of the
energy conservation for a fixed emission angle. We give kinematical
and geometrical explanations of these two solutions. In Sect. VI we
make a theoretical analysis of the angular distribution and find
analytically the boundaries of the angular intervals. We get a
formula, Eq. (\ref{singularity}), for the integral around the
divergent angles. In Sect. VII we present results for the phase-space
function with the new integration method.  In Sect. VIII we discuss how 
this formalism can be applied 
to the 2p-2h response functions of electron and neutrino
scattering.
Finally in Sect. IX we present our conclusions. 

\section{2p-2h response functions}

When considering a lepton that scatters off a nucleus transferring
four-momentum $Q^{\mu}=(\omega,\nq)$, with $\omega$ the energy
transfer and $\nq$ the momentum transfer, one is involved with the 
hadronic tensor
\begin{equation}
W^{\mu\nu}=\sum_f
\langle \Psi_f |J^{\mu}(Q)|\Psi_i\rangle^*
\langle \Psi_f |J^{\nu}(Q)|\Psi_i\rangle
\delta(E_i+\omega-E_f) \,,
\end{equation}
where $J^{\mu}(Q)$ is the electroweak nuclear current operator.

In this paper we take the initial nuclear state as the RFG 
model ground state, $|\Psi_i\rangle=|F\rangle$, with all
states with momenta below the Fermi momentum $k_F$ occupied. The sum
over final states can be decomposed as the sum of one-particle
one-hole (1p-1h) plus two-particle two-hole (2p-2h) excitations plus
additional channels. 
\begin{equation}
W^{\mu\nu} = W^{\mu\nu}_{1p1h} + W^{\mu\nu}_{2p2h} + \cdots
\end{equation}

In the impulse approximation the 1p-1h channel gives the well-known
response functions of the RFG. Notice that MEC also contribute to
these 1p-1h responses; however, here we focus on the 2p-2h channel
where the final states are of the type
\begin{eqnarray}
|\Psi_f\rangle
&=& |1',2',1^{-1},2^{-1}\rangle
\\
|i'\rangle  &=& |\np'_i s'_i t'_i \rangle
\\
|i\rangle   &=& |\nh_i s_i t_i \rangle, 
\kern 1cm 
i,i'=1,2 \,,
\end{eqnarray}
where $\np'_i$ are momenta of relativistic final
nucleons above the Fermi sea, $p'_i>k_F$, with four-momenta
$P'_i=(E'_i,\np'_i)$, and $H_i=(E_i,\nh_i)$ are the four-momenta of
the hole states with $h_i<k_F$. The spin indices are $s'_i$ and
$s_i$, and the isospin is $t_i, t'_i$.

In this paper we study the 2p-2h channel in a fully relativistic framework.
The corresponding hadronic tensor is given by
\begin{eqnarray}
W^{\mu\nu}_{2p-2h}
&=&
\frac{V}{(2\pi)^9}\int
d^3p'_1
d^3p'_2
d^3h_1
d^3h_2
\frac{m_N^4}{E_1E_2E'_1E'_2}
\nonumber \\ 
&&
r^{\mu\nu}(\np'_1,\np'_2,\nh_1,\nh_2)
\delta(E'_1+E'_2-E_1-E_2-\omega)
\nonumber\\
&&
\Theta(p'_1,p'_2,h_1,h_2)
\delta(\np'_1+\np'_2-\nh_1-\nh_2-\nq) \,,
\nonumber\\
\label{hadronic}
\end{eqnarray}
where $m_N$ is the nucleon mass, $V$ is the volume of the system and
we have defined the product of step functions
\begin{equation}
\Theta(p'_1,p'_2,h_1,h_2)=
\theta(p'_2-k_F)
\theta(p'_1-k_F)
\theta(k_F-h_1)
\theta(k_F-h_2) \,.
\end{equation}
The function  $r^{\mu\nu}(\np'_1,\np'_2,\nh_1,\nh_2)$ is the
hadronic tensor for the elementary transition of a nucleon pair 
with the given initial and final momenta, summed up over spin and isospin,
given schematically as
\begin{equation}
r^{\mu\nu}(\np'_1,\np'_2,\nh_1,\nh_2)= \frac{1}{4}\sum_{s,t}
j^{\mu}(1',2',1,2)^*_A
j^{\nu}(1',2',1,2)_A\, ,
\label{elementary}
\end{equation}
which we write in terms of the anti-symmetrized two-body current 
matrix element
$j^{\mu}(1',2',1,2)_A$, to be specified. The factor $1/4$ accounts for
the antisymmetry of the 2p-2h wave function.  Finally, note that the
2p-2h response is proportional to $V$ which is
related to the number of protons or neutrons $Z=N=A/2$
by $V=3\pi^2 Z/k_F^3$. In this work we only consider nuclear targets with pure isospin zero.

In the case of electrons the cross section can be written as a linear 
combination of the longitudinal and transverse response functions defined by
\begin{eqnarray}
R_L&=& W^{00}   \label{rl}\\
R_T&=& W^{11}+W^{22} \, , \label{rt}
\end{eqnarray}
whereas additional response functions arise for neutrino scattering, due to the presence of the axial current.
The generic results coming from the phase-space obtained here
are applicable to all of the response functions.

Integrating over $\bf p'_2$ using the momentum delta function,
Eq. (\ref{hadronic}) becomes a 9D integral
\begin{eqnarray}
W^{\mu\nu}_{2p-2h}
&=&
\frac{V}{(2\pi)^9}\int
d^3p'_1
d^3h_1
d^3h_2
\frac{m_N^4}{E_1E_2E'_1E'_2}
\nonumber \\ 
&&
r^{\mu\nu}(\np'_1,\np'_2,\nh_1,\nh_2)
\delta(E'_1+E'_2-E_1-E_2-\omega)
\nonumber\\
&&
\Theta(p'_1,p'_2,h_1,h_2) \,,
\label{hadronic2}
\end{eqnarray}
where $\bf p'_2= h_1+h_2+q-p'_1$.  After choosing the $\bf q$
direction along the $z$-axis, there is a global rotation symmetry over
one of the azimuthal angles. We choose $\phi'_1=0$ and multiply by 
a factor $2\pi$.  Furthermore, the energy delta function enables 
analytical integration over $p'_1$, 
and so the integral is reduced to 7 dimensions.
In general the calculation has to be done numerically. 
Under some approximations~\cite{Donnelly:1978xa,Van80,Alb84,Gil97}
the number of dimensions can be further reduced,
but this cannot be done in the fully relativistic calculation.

In this paper we study different methods to evaluate the above
integral numerically and compare the relativistic and the
non-relativistic cases.  In the non-relativistic case we reduce the
hadronic tensor to a 2D integral.  This can be done when the function
$r^{\mu\nu}$ only depends on the differences $\nk_i=\np'_i-\nh_i$,
$i=1,2$.

As we want to concentrate on the numerical procedure without further
complications derived from the momentum dependence of the currents, in
this paper we start by setting the elementary function to a constant
$r^{\mu\nu}=1$.  Hence, we focus on the genuine kinematical effects
coming from the two-particle-two-hole phase-space alone. In particular,
the kinematical relativistic effects arising from the energy-momentum
relation are contained in the energy conservation delta function that
determines the analytical behavior of the hadronic tensor, 
where the energy-momentum relation is $E=\sqrt{k^2+m_N^2}$,
 and in the
Lorentz contraction coefficients $m_N/E_i$.  Obviously the results
obtained here for constant $r^{\mu\nu}$ 
will be modified when including the two-body physical
current. But as the final result is model-dependent, it is not
possible to disentangle whether the differences found are due to the 
current model employed or to the approximations
(relativistic or not) used to perform the numerical evaluation of the
integral.  In fact all of the models of 2p-2h response functions should
agree at the level of the 2p-2h phase-space integral $F(q,\omega)$ defined
as
\begin{eqnarray}
F(q,\omega)
&\equiv&
\int
d^3p'_1
d^3h_1
d^3h_2
\frac{m_N^4}{E_1E_2E'_1E'_2}
\nonumber \\ 
&&
\delta(E'_1+E'_2-E_1-E_2-\omega)
\Theta(p'_1,p'_2,h_1,h_2) \,,
\nonumber\\
\label{phase}
\end{eqnarray}
with $\bf p'_2= h_1+h_2+q-p'_1$. Calculation of this function should
be a good starting point to compare and congenialize different nuclear 
models.

\section{Non-relativistic 2p-2h phase-space}

\subsection{Semi-analytical integration}

First we recall the semi-analytical method of~\cite{Van80} that
was used later in  \cite{Alb84,DePace03}, for instance, to
compute the non-relativistic 2p-2h transverse response function in
electron scattering. We shall use this method to check the
numerical 7D quadrature both in the relativistic and non-relativistic cases. 

 We start with the 12D expression
for the phase-space function Eq. (\ref{hadronic})
\begin{eqnarray}
F(q,\omega)
&=&
\int
d^3p'_1
d^3p'_2
d^3h_1
d^3h_2
\nonumber \\ 
&&
\delta(E'_1+E'_2-E_1-E_2-\omega)
\nonumber\\
&&
\Theta(p'_1,p'_2,h_1,h_2)
\delta(\np'_1+\np'_2-\nh_1-\nh_2-\nq) \,.
\nonumber\\
\label{phase2}
\end{eqnarray}
The procedure is first to perform the integral over energy.
Following~\cite{Van80} we change variables
\begin{eqnarray}
\nl_1= \frac{\np'_1-\nh_1}{k_F}
&&
\nl_2= \frac{\np'_2-\nh_2}{k_F}
\label{variables1}\\
\nx_1= \frac{\np'_1+\nh_1}{2 k_F}
&&
\nx_2= \frac{\np'_2+\nh_2}{2 k_F} \,.
\label{variables2}
\end{eqnarray}
We also define the following non-dimensional variables
\begin{eqnarray}
\nq_F &\equiv& \frac{\nq}{k_F} \\
\nu &\equiv& \frac{m_N\omega}{k_F^2} \,.
\end{eqnarray}
In terms of these variables  the 2p-2h phase-space function is
\begin{eqnarray}
F(q,\omega) &=&
(2\pi)^2k_F^7 m_N
\int
\frac{d^3l_1}{l_1^3}
\frac{d^3l_2}{l_2^3}
\nonumber\\
&&
\delta(\nl_1+\nl_2-\nq_F)A(l_1,l_2,\nu)\,,
\end{eqnarray}
where we use the Van Orden function defined as
\begin{eqnarray}
A(l_1,l_2,\nu)
&=&
\frac{l_1^3l_2^3}{(2\pi)^2}
\int d^3x_1d^3x_2
\delta(\nu-\nl_1\cdot\nx_1-\nl_2\cdot\nx_2)
\nonumber\\
&&
\theta\left( 1-\left|\nx_1-\frac{\nl_1}{2}\right| \right)
\theta\left( 1-\left|\nx_2-\frac{\nl_2}{2}\right| \right)
\nonumber\\
&&
\theta\left( \left|\nx_1+\frac{\nl_1}{2}\right| -1 \right)
\theta\left( \left|\nx_2+\frac{\nl_2}{2}\right| -1 \right) \, . \nonumber \\
&&
\end{eqnarray}
This function was computed analytically in \cite{Van80}. In this
work we have checked that expression because we found 
a typo in one of the terms in the original reference
(that typographical error does not affect the
results of the cited reference). We give in the Appendix the
correct result for future reference.

Integrating now over the momentum $\nl_2$ we get
\begin{eqnarray}
F(q,\omega) &=&
(2\pi)^2k_F^7 m_N
\int
\frac{d^3l_1}{l_1^3|\nq_F-\nl_1|^3}
\nonumber\\
&&
A(l_1,|\nq_F-\nl_1|,\nu) \,.
\end{eqnarray}
The integral over the azimuthal angle $\phi_1$ of $\nl_1$ gives $2\pi$.
Finally changing to the variables 
\begin{equation}\label{variables3}
x= l_1, \kern 1cm
y=  |\nq_F-\nl_1|
\end{equation}
we obtain
\begin{eqnarray}
F(q,\omega) &=&
(2\pi)^3\frac{k_F^7 m_N}{q_F}
\int_0^{x_{\rm max}}
\frac{dx}{x^2}
\int_{|q_F-x|}^{q_F+x}
\frac{dy}{y^2}
 A(x,y,\nu) \,,
\nonumber\\
\label{fasico2D}
\end{eqnarray}
where the maximum value of $x$ (or $k_1/k_F$) is obtained from
the energy conservation and momentum step functions included 
implicitly in the function $A(x,y,\nu)$. In the appendix
we derive the inequality 
\begin{equation}
x \leq x_{\rm max} \equiv 1 + \sqrt{2(1+\nu)}.
\end{equation}

The 2D integral over the variables $x,y$ has to be performed
numerically.

\subsection{Numerical integration}

The simplicity of the Fermi gas model used in this paper allows us to
compute the 2p-2h hadronic tensor as a 7D dimensional integral as
shown below. Note that in a more sophisticated model where the nuclear
distribution details are taken into account, like shell models or the
spectral function-based models, some of the numerical problems linked
to the particular Jacobian appearing here and in the following
section, can be avoided, at the price of increasing the number of
integrals or sums over shell-model states, thus making the
calculations harder. The local Fermi gas used by Nieves {\it et al.,} is really an
average of different Fermi gases at different densities, but the basic
Fermi gas equations are the same as here.

The hadronic tensor for the elementary 2p-2h transition,
Eq.~(\ref{elementary}), contains the direct and exchange matrix
elements of the two-body current operator. If one neglects the
interference between the direct and exchange terms it is possible to
express $r^{\mu\nu}$ as a function of $x,y$ only, and one can use the
formalism of the above section to reduce the calculation of the 2p-2h
hadronic tensor to a 2D integral.  In the general case the
interference cannot be neglected.  It is then necessary to evaluate a
7D integral numerically.  Thus, in this work we also
compute the phase-space function, Eq.~(\ref{phase}), numerically.  This
will allow us firstly to check the numerical procedures by comparison
with the semi-analytical method of the previous section, secondly
to determine the number of integration points needed to obtain
accurate results and thirdly to optimize the computational effort.
This numerical study will be very useful when including
actual nuclear currents.
 
Starting with Eq.~(\ref{phase}), we compute the
integrand for $\phi'_1=0$ (the azimuthal angle of $\np'_1$), and
multiply by $2\pi$.  Then we use the $\delta$ of energies to integrate
over the variable $p'_1$, for fixed momenta $\nh_1,\nh_2$ and emission
angle $\theta'_1$.  To do so we first define the total momentum of the
two particles that is fixed by momentum conservation
\begin{equation}
\np'= \np'_1+\np'_2 = \nh_1+\nh_2+\nq.
\end{equation}
We then change from variable $p'_1$ to variable $E'$:
\begin{equation}
E'  = E'_1+E'_2  = 
\frac{p'_1{}^2}{2m_N}
+\frac{(\np'-\np'_1)^2}{2m_N} \,.
\label{eprima}
\end{equation}
By differentiation
with respect to $p'_1$, we obtain 
\begin{equation}
\left|\frac{dp'_1}{dE'}\right|
=\frac{m_N}
{\left| p'_1-\np'_2\cdot\hp'_1 \right|} \,,
\label{jacobiano1}
\end{equation}
where $\hp'_1 = \np'_1/p'_1$ is the unit vector in the direction of
the first particle. 
Integrating now over $E'$, energy conservation is obtained as
\begin{equation}
E'=E_1+E_2+\omega \,.
\end{equation}
Substituting Eq.~(\ref{eprima})
a second degree equation is obtained for $p'_1$ 
\begin{equation}
2p'_1{}^2+p'{}^2-2\np'\cdot\np'_1= 2m_NE' \,.
\end{equation}
So we see that there can be two values of the nucleon momentum 
compatible with energy conservation, for fixed emission angle.
We denote the two solutions by
\begin{equation}
p'_1{}^{(\pm)}=\frac12\left[
v \pm \sqrt{ v^2-4\left(\frac{p'{}^2}{2}-m_NE'\right)}\phantom{|}
\right] \,,
\label{momentum}
\end{equation}
where we have defined
\begin{equation}
v\equiv \np'\cdot\hp'_1 \,.
\end{equation}
Using this result we finally evaluate the phase-space function as
the 7D integral
\begin{eqnarray}
F(q,\omega)
&=&
2\pi
\int
d^3h_1
d^3h_2
d\cos\theta'_1
\label{integral7D}\\ 
&&
\sum_{\alpha=\pm}
\left.
\frac{p'_1{}^2 m_N}
{\left| p'_1-\np'_2\cdot\hp'_1 \right|}
\Theta(p'_1,p'_2,h_1,h_2)
\right|_{p'_1= p'_1{}^{(\alpha)}} \,,
\nonumber
\end{eqnarray}
where the sum inside the integral runs over the two solutions
$p'_1{}^{(\pm)}$ of the energy conservation equation.

\subsection{Asymptotic expansion}

It is of interest to quote the limit $\omega \rightarrow \infty$,
because it can also be used for testing the numerical integration.
The most useful case applies for $k_F,q \ll \omega$, when one can
neglect all momenta compared with the energy transfer $\omega$,
because the phase-space integral can be performed analytically.  Note
that for the scattering reactions of interest this limit is not
physical (because $\omega< q$, namely spacelike, for real
particles). It is only a mathematical property of the function $F$,
that is well defined for all the $\omega$ values, not only the
physical ones.  We start writing the momentum of the first particle,
Eq.~(\ref{momentum}), as
\begin{equation}
p'_1 = \frac{v}{2} \pm\frac12\sqrt{D}
\end{equation}
with the discriminant
\begin{equation}
D = v^2 - 2p'^2 + 4m_NE' \,.
\end{equation}
The limit $\omega\rightarrow\infty$ can be obtained by  noticing 
that $v$ and $p'$
do not depend on $\omega$, but only on the momenta $\nh_1, \nh_2, q$,
and that $E'= E_1+E_2+\omega \sim \omega$. Then 
\begin{equation}
D \sim 4m_N\omega
\end{equation}
and  the positive solution for the momentum is
\begin{equation}
p'_1 \sim \sqrt{m_N\omega} \,.
\end{equation}
That is, each nucleon exits the nucleus taking half of the available energy.

On the other hand, using (\ref{momentum}),  we note that 
the denominator in Eq.~(\ref{integral7D}) can be written
as
\begin{equation}
p'_1-\np'_2\cdot\hp'_1 = \pm \sqrt{D} \sim \pm 2\sqrt{m_N\omega} \,.
\end{equation}
Then
\begin{eqnarray}
F(q,\omega) 
& \stackrel{\omega\rightarrow\infty}{\longrightarrow} &
F_{a}(q,\omega)
\nonumber\\
&\equiv&
2\pi \int d^3h_1 d^3h_2 d\cos\theta'_1 \frac{m_N}{2}\sqrt{m_N\omega}
\nonumber\\
&=& 
4\pi
\left(\frac43 \pi k_F^3 \right)^2
\frac{m_N}{2}\sqrt{m_N\omega} \,.
\label{asymptotic}
\end{eqnarray}

Thus, for high energy, the non-relativistic phase-space function increases as
$\sqrt{\omega}$. We shall see in the next section a different behavior in 
the relativistic case.

\subsection{Non-relativistic results}

\begin{figure}
\includegraphics[width=8cm, bb=170 280 450 780]{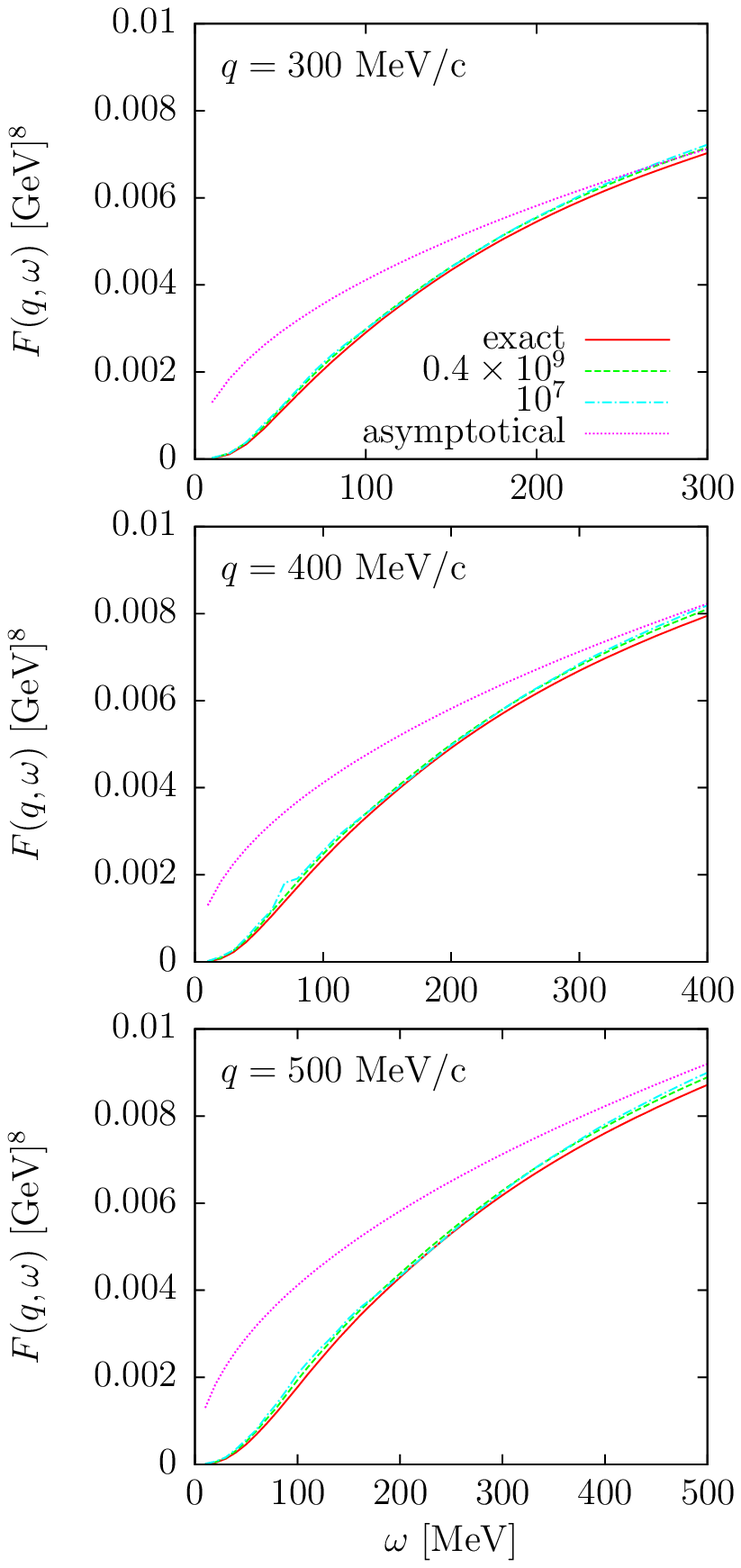}
\caption{
\label{fig4} 
(Color online) 
Non-relativistic phase-space function calculated 
for $\omega = 300, 400, 500$ MeV, using a numerical and a
semi-analytical approach.  The number of points used in two numerical
integrations are indicated in the plot.  We also show the asymptotic
function for comparison. }
\end{figure}

In Fig.~\ref{fig4} we show the non-relativistic phase-space function
$F(q,\omega)$ as a function of $\omega$ for three typical values of
the momentum transfer $q=300, 400$ and 500 MeV/c. The Fermi momentum
is $k_F= 225$ MeV/c. We compare the two computational methods: the
semi-analytical of Eq.~(\ref{fasico2D}) and the numerical 7D
integration of Eq.~(\ref{integral7D}). The semi-analytical result is
essentially exact, because we can choose a very small integration step
for the 2D integral (using steps of 0.02 or 0.01 the results do not
change in the scale of the figure). However the 7D integral is
computationally time-consuming and the integration step cannot be very
small.  Here we compute the integral with a ``straightforward''
method, as an average over a grid with $n$ total integration points,
uniformly distributed.  For large $n$ the straightforward integration
should give results similar to the Montecarlo methods used in previous
calculations~\cite{Van80,DePace03}.  The number of points chosen for
this calculation was 25 for the variable $\theta'_1$ (although it can
safely be reduced to 16) and 16 for each one of the remaining
dimensions. In total the number of points is $n= 0.42\times 10^9$.
This is well above the maximum number $n=10^6$--$10^7$, typical of
previous calculations~\cite{Van80,DePace03} performed using Monte
Carlo techniques.  Using 10 integration points in each dimension gives
very similar results, except for some $\omega$ regions where the
numerical error is manifested in an apparently slightly less smooth
behavior. Increasing the number of points would improve the results;
however, this is not practical because the inclusion of the two-body
current would make the calculation too slow.  The semi-analytical and
numerical results are quite similar, the difference between them being
of a few percent.  For comparison we also show the asymptotic limit
$\omega\rightarrow\infty$, computed using the analytical expression in
Eq.~(\ref{asymptotic}), which is proportional to $\sqrt{\omega}$. We
see that for high $\omega$ the function $F(q,\omega)$ becomes close to
the asymptotic value $F_{a}(q,\omega)$.  For $q=300$ MeV/c the
asymptotic value is almost reached at the photon point
$\omega=q$. When $q$ increases, so does the distance to the asymptote
at the photon point.

\section{Relativistic 2p-2h phase-space}

Having two independent calculations 
of the phase-space function $F(q,\omega)$  in the non-relativistic 
limit, we now consider the case of the fully relativistic calculation
as given by Eq.~(\ref{phase}). This involves adding the 
Lorentz-contraction  $m_N/E$ factors and using relativistic kinematics in the
energy $\delta$-function. Following the scheme of the previous section, 
again  azimuthal symmetry allows one to fix $\phi'=0$ and multiply by $2\pi$.
To integrate over $p'_1$ we change to the variable 
\begin{equation}
E'=E'_1+E'_2=\sqrt{p'_1{}^2+m_N^2}+\sqrt{(\np'-\np'_1)^2+m_N^2} \,,
\end{equation}
where again $\np'=\nh_1+\nh_2+\nq$ is the final momentum for a fixed
pair of holes.
By differentiation we arrive at the following Jacobian:
\begin{equation}
\left|\frac{dp'_1}{dE'}\right|
=\left| \frac{p'_1}{E'_1}-\frac{\np'_2\cdot\hp'_1}{E'_2} \right|^{-1} \,.
\label{jacobiano2}
\end{equation}
The non-relativistic Jacobian of
Eq.~(\ref{jacobiano1}) is recovered for low energies $E'_1\simeq
E'_2\simeq m_N$.
As before, integration over $E'$ gives $E'=E_1+E_2+\omega$
and the phase-space function becomes 
\begin{eqnarray}
F(q,\omega)
&=&
2\pi 
\int
d^3h_1
d^3h_2
d\theta'_1\sin\theta'_1
\frac{m_N^4}{E_1E_2}
\label{integral7Drel}\\ 
&\times &
\sum_{\alpha=\pm}
\left.
\frac{p'_1{}^2}
{\left| \frac{p'_1}{E'_1}-\frac{\np'_2\cdot\hp'_1}{E'_2} \right|}
\frac{ \Theta(p'_1,p'_2,h_1,h_2) }{ E'_1E'_2 }
\right|_{p'_1= p'_1{}^{(\alpha)}} \,,
\nonumber
\end{eqnarray}
where again the sum inside the integral runs over the two solutions
$p'_1{}^{(\pm)}$ of the energy conservation equation
\begin{equation}  \label{segundo-grado}
p'_1{}^{(\pm)}=
\frac{1}{\tilde{b}}
\left[
\tilde{a}\tilde{v}
\pm
\sqrt{\tilde{a}^2-\tilde{b}m_N^2}
\right] \,.
\end{equation}
The definitions of the quantities $\ta,\tb,\tv$ are given in the
Appendix.  Note that there is a difference between our Jacobian in
Eq. (\ref{integral7Drel}) and that given in Eqs. (15--17) of 
\cite{Lal12}).  

The relativistic approach is more involved than
the non-relativistic one because it requires taking the square twice
in the original equation to eliminate the squared roots in the
energies. This can introduce spurious solutions for $p'_1$ depending
on the kinematics, that have to be eliminated from the above sum in
the numerical procedure.  This is not a trivial task and details are
provided in the Appendix. The appearance of spurious solutions is a
difference between the relativistic and non-relativistic methods.  A
second one will be discussed below in relation to a divergence of the
integrand. Therefore the relativistic calculation is very involved and
it cannot be derived by simply extending the non-relativistic code.
We devote the rest of this section to explain in detail how to get the
fully relativistic answers.

\subsection{Relativistic asymptotic expansion}

Although it is not possible to derive a semi-analytical expression for
$F(q,\omega)$ as in the non-relativistic case, it is still possible to
take the limit $\omega\rightarrow \infty$ and obtain an analytical
result. As in the non-relativistic case, we assume $k_F,q \ll
\omega$.  If we add the condition $m_N\ll\omega$, we can neglect the
momenta and energies of the two holes and write
\begin{equation}
E'\sim\omega   \kern 1cm \np'\sim\nq \,.
\end{equation}
We can also compute the quantities with tildes that appear in the
solution of the energy conservation (see Appendix), obtaining  
\begin{eqnarray}
\ta & \sim & \frac{\omega}{2} \\
\tv & \sim & \frac{\nq\cdot\hp'_1}{2\omega} \\
\tb & \sim & 1 \,.
\end{eqnarray}
Then the discriminant of Eq.~(\ref{segundo-grado}) becomes
\begin{equation}
\ta^2 - \tb m_N^2 \sim \frac{\omega^2}{4} - m_N^2 \sim \frac{\omega^2}{4} \,.
\end{equation}
Therefore the allowed solution of the energy conservation equation is
\begin{equation}\label{asintotico1}
p'_1 \sim \frac{\nq\cdot\hp'_1}{4} + \frac{\omega}{2} \sim \frac{\omega}{2}\,.
\end{equation}
Thus in this limit each nucleon carries half the total energy and momentum 
\begin{equation} \label{asintotico2}
E'_1 \sim  p'_1 \sim E'_2 \sim p'_2 \sim \frac{\omega}{2}\,.
\end{equation}
Now the Jacobian, the denominator in Eq.~(\ref{integral7Drel}), can be computed as
\begin{equation} \label{asintotico3}
d \equiv 
 \frac{p'_1}{E'_1}-\frac{\np'_2\cdot\hp'_1}{E'_2} 
=
1-\frac{(\np'-\np'_1)\cdot\hp'_1}{E'_2} 
\sim 1+\frac{p'_1}{E'_2}
\sim 2 \,.
\end{equation}
Collecting Eqs. (\ref{asintotico1},\ref{asintotico2},\ref{asintotico3}),
the integrand in Eq.~(\ref{integral7Drel}) becomes
\begin{equation}
\frac{p'_1{}^2}{d} 
\frac{m_N^4}{E_1E_2E'_1E'_2}
\sim
\frac{\omega^2}{8}
\frac{m_N^4}{m_N^2\omega^2/4}
= \frac{m_N^2}{2} \,.
\end{equation}
Finally, performing the integral we obtain the following asymptotic expression
\begin{equation}
F(q,\omega) 
\stackrel{\omega\rightarrow\infty}{\longrightarrow}
F_{a}(q,\omega)
= 
4\pi
\left(\frac43 \pi k_F^3 \right)^2
\frac{m^2_N}{2}.
\label{asymptotic2}
\end{equation}
In contrast with the non-relativistic behavior, that increases 
monotonically as $\sqrt{\omega}$, 
the relativistic result~(\ref{asymptotic}) goes to a constant. 
The Lorentz contraction factors  $E/m_N$  balance the $\omega^2$ 
behavior coming from the phase-space.  
This analytical result for high $\omega$ will be useful for comparison
of  the numerical results for high $\omega$.

\subsection{Relativistic straightforward calculation}

\begin{figure}
\includegraphics[width=8cm, bb=170 280 450 780]{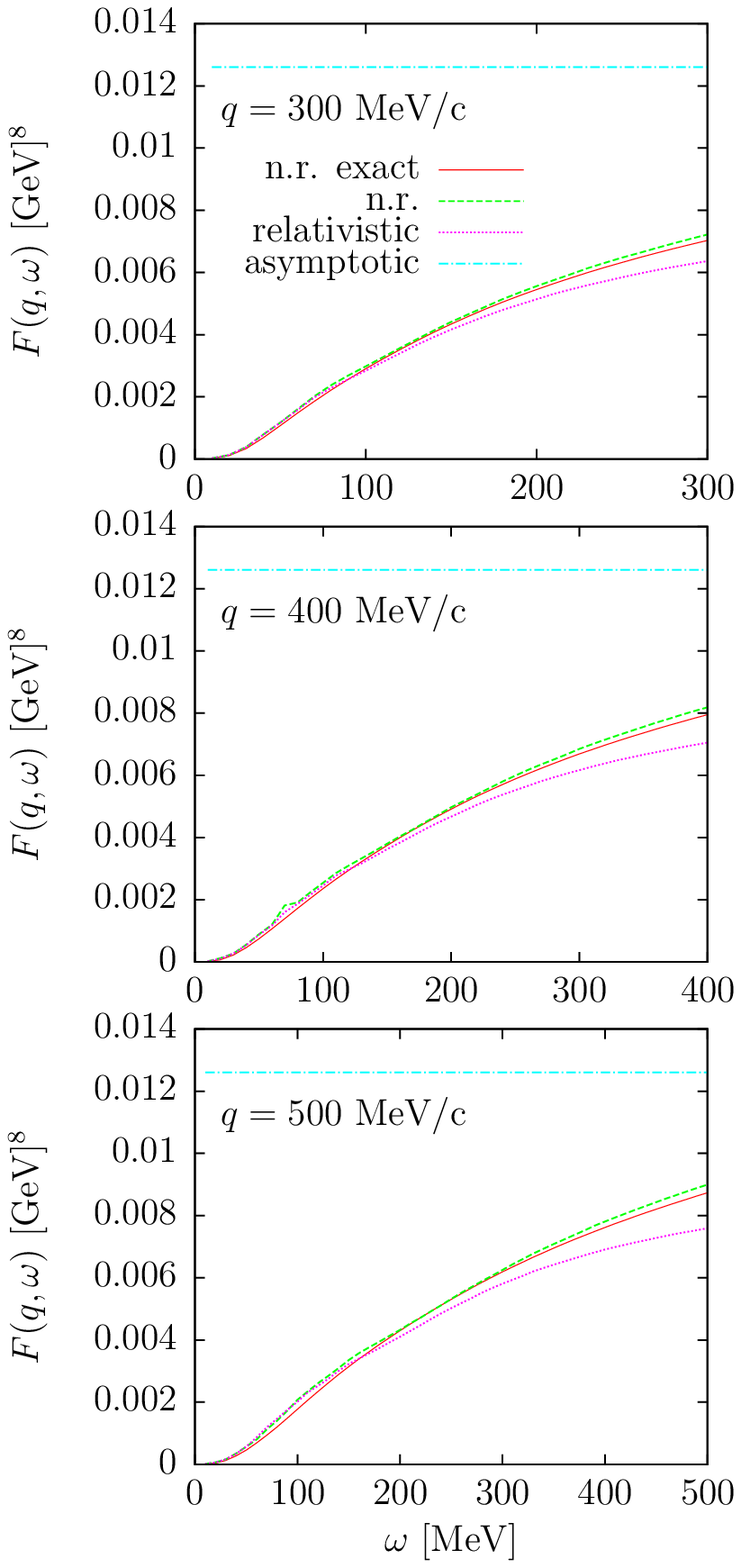}
\caption{
\label{fig5} 
(Color online) 
Relativistic phase-space function
for $q = 300, 400, 500$,  calculated  using
straightforward integration, compared with the non-relativistic 
calculation using the
semi-analytical approach.  }
\end{figure}

\begin{figure}
\includegraphics[width=8cm, bb=170 280 450 780]{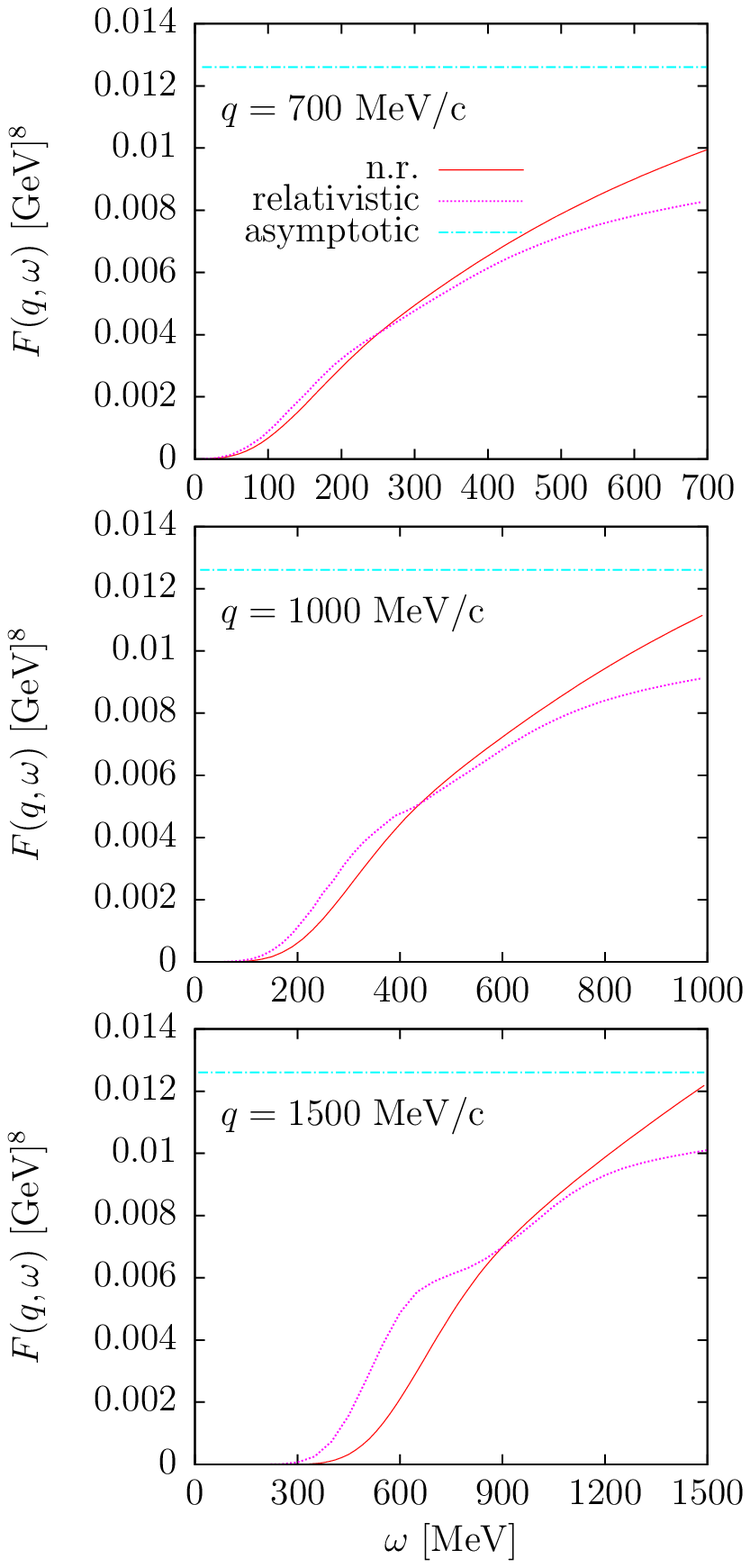}
\caption{
\label{fig6} 
(Color online) 
Relativistic phase-space function
for $q = 700, 1000, 1500$, calculated using 
straightforward integration  
compared with the non-relativistic calculation using the semi-analytical
approach.  }
\end{figure}

Before going to the high-$q$ region, we first check the relativistic
phase-space function results by comparison with the non-relativistic
counterpart. Both should agree for low energy.  We proceed by
performing a straightforward numerical integration of
Eq.~(\ref{integral7Drel}) as in the non-relativistic case.  In
Fig.~\ref{fig5} we show the results of this comparison for $q=300,
400$ and $500$ MeV/c. We also show both the numerical and ``exact''
({\it i.e.} using the semi-analytical formula) non-relativistic function
$F(q,\omega)$. A uniform distribution with 10 points for each
dimension is employed in the 7D integrations. As expected,
relativistic and non-relativistic results agree at low energy. The
relativistic effects consist of a reduction of the strength at high
energy.  The amount of this reduction is very small for $q=300$ MeV/c,
where the non-relativistic approximation can be safely used, and
increases with $q$, reaching about $15\%$ for $q=500$ MeV/c.  Thus for
low $q$ we agree that a number of $\sim 10^7$ points is adequate for
numerical integration purposes.  In Fig.~5 the asymptotic limit
$F_{a}(q,\omega)$ of the relativistic phase-space,
Eq.~(\ref{asymptotic2}), is also shown. For these low $q$ values,
$F(q,\omega)$ is still far below the asymptote.

Larger relativistic effects are expected for intermediate to large
momentum transfer. In Fig.~\ref{fig6} we display $F(q,\omega)$ for
$q=700, 1000$ and 1500 MeV/c compared with the exact non-relativistic
results. Using straightforward 7D integration we need to increase the
number of points to 16 for each dimension in order to reach some
stability of the results shown in Fig.~\ref{fig6}.  However, we find
that full convergence would need more points. In fact for $q=700$
MeV/c a small deviation with respect to the exact result can be noticed at
low $\omega$.  This deviation increases with $q$ and turns into a
prominent structure with a ``shoulder'' shape for $q=1500$ MeV/c. One
could be tempted to attribute this effect to relativity. But this is
not the case because the same behavior is also present in a
non-relativistic numerical calculation. As we will explain below, this
is just a consequence of the inadequacy of the straightforward integration 
method at high $q$.
This problem affects only the 
inner integral over $\theta'_1$.  Below we address this issue by a
detailed analysis of the $\theta'_1$ dependence of the integrand.


\section{Angular distribution}


\subsection{Frozen phase-space function}

We start fixing a value of $q=3$ GeV/c that is high enough to amplify
the misbehavior found above, and also allows  
to simplify the analysis that follows. 
In fact  we note that for very high $q \gg k_F$
all of the hole momenta $\nh_1$, $\nh_2$ could safely be neglected 
inside the integral as a first approximation. Since this implies that the 
initial particles are at rest, we denote this limit 
the ``frozen nucleon approximation``.
 In particular, 
the energies of the holes can be substituted by the nucleon mass
in the $\delta$ function:
\begin{eqnarray}
F(q,\omega) 
&\sim& 
\int  d^3h_1  d^3h_2  d^3p'_1 
\delta(E'_1+E'_2-\omega-2m_N) 
\nonumber\\
&&\times \Theta(p'_1,p'_2,0,0)\frac{m_N^2}{E'_1E'_2} \,,
\end{eqnarray}
where $\np'_2=\nq-\np'_1$.
Because the integrand does not depend on the hole momenta, one can
directly integrate out those variables
\begin{eqnarray}
F(q,\omega) 
&\sim &
\left(\frac{4}{3}\pi k_F^3\right)^2
\int d^3p'_1 
\delta(E'_1+E'_2-\omega-2m_N) 
\nonumber\\
&&\times \Theta(p'_1,p'_2,0,0)\frac{m_N^2}{E'_1E'_2} \,.
\end{eqnarray}
Now the integral over $p'_1$ can be done analytically as before using
the delta function, with the same Jacobian evaluated for $h_1=h_2=0$.
The integral over $\phi'_1$ gives again a factor $2\pi$.
\begin{eqnarray}
F(q,\omega)
&\sim&
2\pi m_N^2
\left(\frac{4}{3}\pi k_F^3\right)^2
\int d\theta'_1\sin\theta'_1
\label{frozen-integral}\\ 
&\times &
\sum_{\alpha=\pm}
\left.
\frac{p'_1{}^2}
{\left| \frac{p'_1}{E'_1}-\frac{\np'_2\cdot\hp'_1}{E'_2} \right|}
\frac{ \Theta(p'_1,p'_2,0,0) }{ E'_1E'_2 }
\right|_{p'_1= p'_1{}^{(\alpha)}} \,.
\nonumber
\end{eqnarray}
Thus in this approximation, the phase-space function is reduced to a
one-dimensional integral over the emission angle $\theta'_1$, which has
to be performed numerically.

The frozen nucleon approximation represents just a particular case of the 
mean-value theorem for the integral over $\nh_1,\nh_2$. We denote with a bar 
the quantities computed by the mean-value theorem. Thus we define the 
barred  phase-space function 
\begin{eqnarray}
\overline{F}(q,\omega) &=& 
\left(\frac{4}{3}\pi k_F^3\right)^2
\int d^3p'_1 
\delta(E'_1+E'_2-\omega-E_1-E_2)
\nonumber\\
&&\times \Theta(p'_1,p'_2,h_1,h_2)
 \frac{m_N^4}{E_1E_2E'_1E'_2} \,,
\label{barred}
\end{eqnarray}
where $\np'_2=\nh_1+\nh_2+\nq-\np'_1$, and ($\nh_1,\nh_2)$ are a pair
of fixed momenta below the Fermi sea.  Going further, we will later
turn to the question of how to choose the average nucleon momenta
$\nh_1,\nh_2$ for low $q$.  For high $q$ we expect
this function not to depend too much on the chosen values. So at this point
we restrict our study to $\overline{F}(q,\omega)$ in the
frozen nucleon approximation, {\it i.e.}, for $h_1=h_2=0$.

\subsection{Numerical analysis}

\begin{figure}
\includegraphics[width=8cm, bb=130 450 410 780]{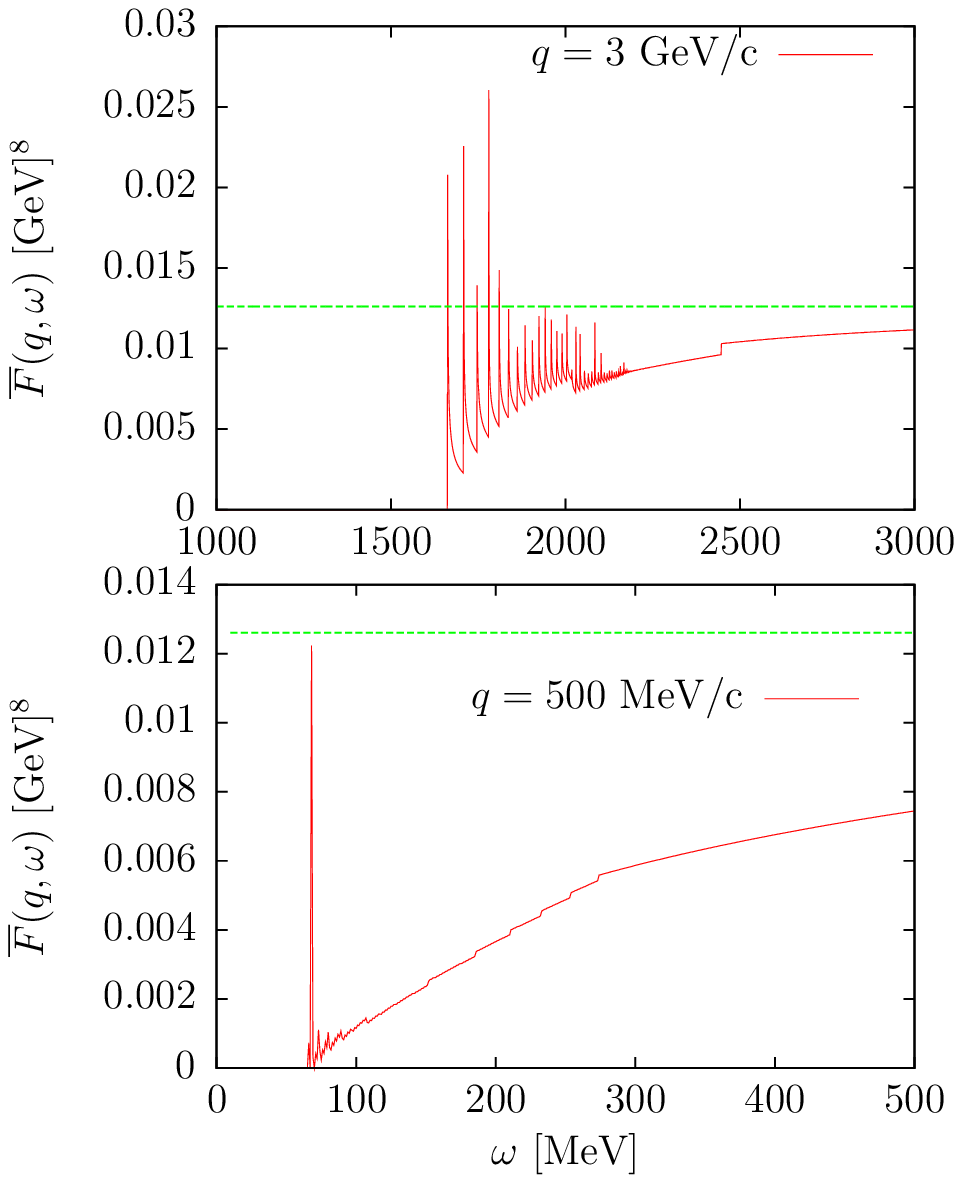}
\caption{
\label{fig8bis} 
(Color online) 
Phase-space function
for $q = 3$ and 0.5 GeV/c, computed using the frozen nucleon approximation
for fixed hole momenta  $h_1=h_2=0$, 
using 100 integration points in emission angle.
}
\end{figure}

We have computed $\overline{F}(q,\omega)$ in the frozen nucleon
approximation using 100 points to perform
the numerical integral over the emission angle $\theta'_1$. Results
are shown in Fig.~\ref{fig8bis}. A misbehavior due to
numerical error is now evident. 

\begin{figure}
\includegraphics[width=8cm, bb=130 280 410 780]{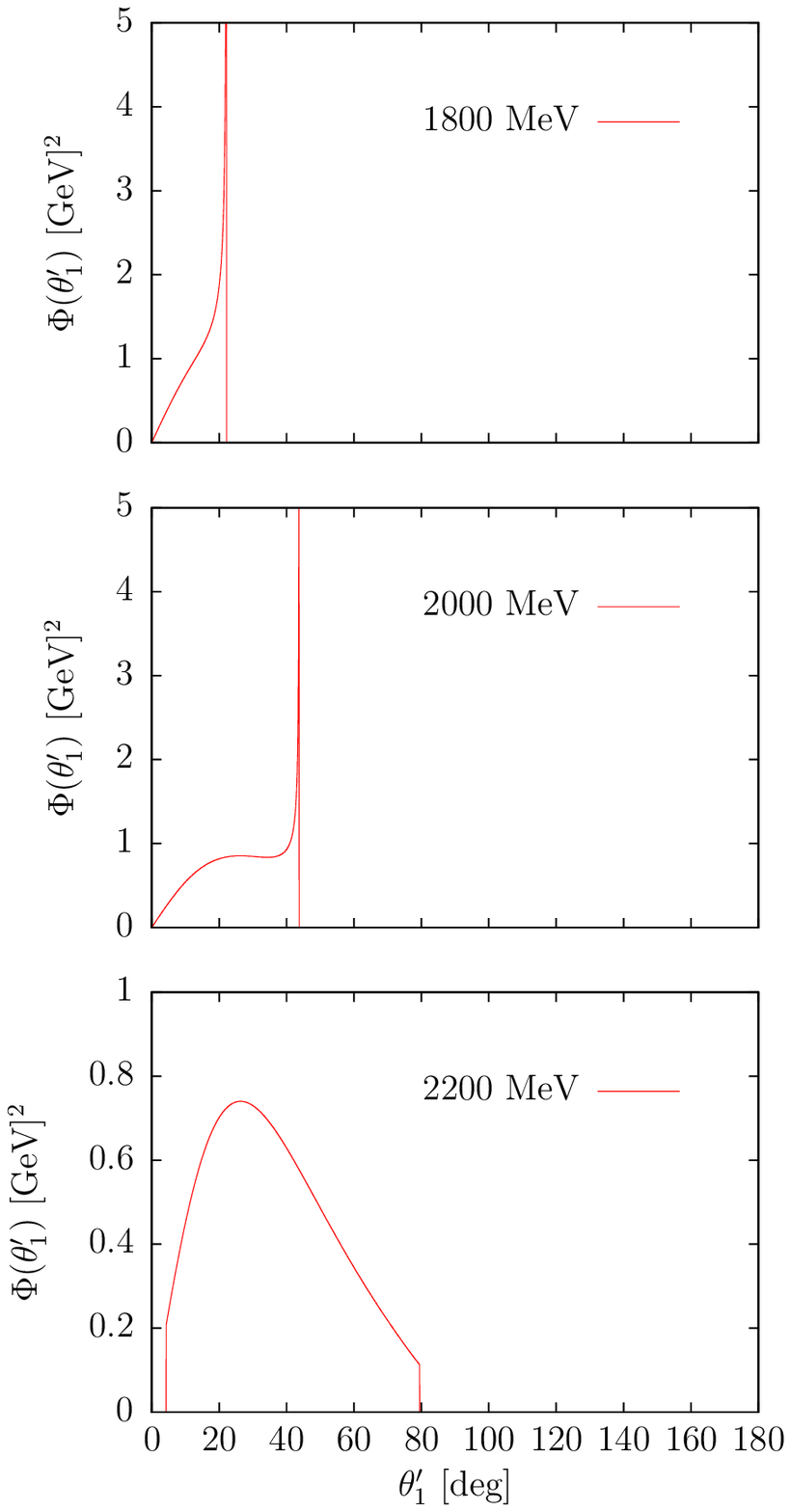}
\caption{
\label{fig10} 
(Color online) 
Angular dependent phase-space function
for $q = 3$ GeV/c, 
for fixed hole momenta  $h_1=h_2=0$, computed for three values of $\omega$
below the quasielastic peak,
as a function of the emission angle $\theta'_1$.
}
\end{figure}

\begin{figure}
\includegraphics[width=8cm, bb=130 600 410 780]{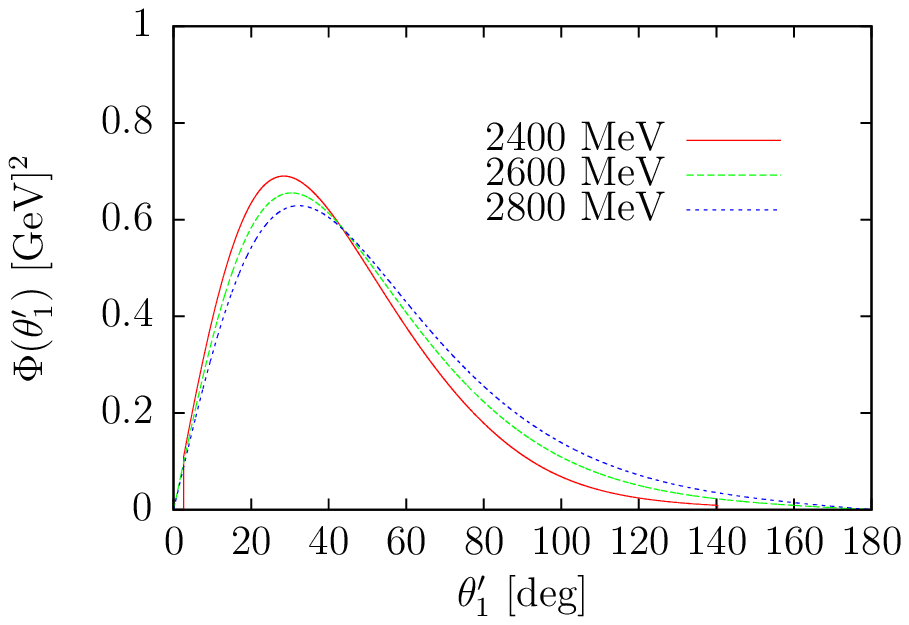}
\caption{
\label{fig11} 
(Color online) 
Angular dependent phase-space function
for $q = 3$ GeV/c, 
for fixed hole momenta  $h_1=h_2=0$, computed for three values of $\omega$
above the quasielastic peak,
as a function of the emission angle $\theta'_1$.
}
\end{figure}

The reason for the appearance of discontinuities by numerical integration 
becomes  apparent by
examining the angular dependence of the integrand.  We define the
angular distribution function, for fixed values of $(q,\omega)$ and
$\nh_1, \nh_2$, as
\begin{eqnarray}
\Phi(\theta'_1)
&=&
 \sin\theta'_1 \int p'_1{}^2 dp'_1 \delta(E_1+E_2+\omega-E'_1-E'_2)
\nonumber\\
&&\times
\Theta(p'_1,p'_2,h_1,h_2)
 \frac{m_N^4}{E_1E_2E'_1E'_2} \,,
\nonumber\\
&=&
\sum_{\alpha=\pm}
\left.
\frac{  m_N^4\sin\theta'_1 p'_1{}^2 \Theta(p'_1,p'_2,h_1,h_2)  }
{  E_1E_2E'_1E'_2 
   \left| \frac{p'_1}{E'_1}-\frac{\np'_2\cdot\hp'_1}{E'_2} \right|}
\right|_{p'_1= p'_1{}^{(\alpha)}} 
\end{eqnarray}
where once more $\np'_2=\nh_1+\nh_2+\nq-\np'_1$, 
such that the phase-space function is obtained by integration 
 over the emission angle $\theta'_1$
\begin{equation} \label{angular}
\overline{F}(q,\omega)=
\left(\frac{4}{3}\pi k_F^3\right)^2
2\pi \int_0^{\pi} d\theta'_1 \Phi(\theta'_1) \,.
\end{equation}
The function $\Phi(\theta'_1)$ thus measures the distribution of final
nucleons as a function of the angle $\theta'_1$.  This function is
computed analytically, given by the integrand in Eq.~(\ref{frozen-integral}). 

Results for $\Phi(\theta'_1)$ are shown in Fig.~\ref{fig10} for
$h_1=h_2=0$, $q=3$ GeV/c and for three values of $\omega=1800$, 2000
and 2200 MeV. For low $\omega$, the function $\Phi(\theta'_1)$ is
different from zero in a narrow angular interval at low angles. At the
upper limit of the interval a divergence appears as a thin peak which is
infinitely high due to a zero in the denominator.  The angular interval
increases with $\omega$ as does the value of the divergent angle. For
$\omega=2200$ MeV there is no divergence because Pauli blocking
forbids reaching the divergent angle. Instead the angular distribution
starts and ends abruptly due to the discontinuity produced by the step
functions.  Note that the values of $\omega$ shown in Fig.~\ref{fig10}
are located below the
quasielastic (QE) peak, that is defined by
\begin{equation}
\omega=\sqrt{m_N^2+q^2} -m_N \,.
\end{equation}
For $q=3$ GeV/c, the QE peak is located roughly at $\omega=2200$ MeV.

The
situation is different for $\omega$ values above the QE peak.  In
Fig.~\ref{fig11} we show in the same plot the angular distribution
for $\omega= 2400$, 2600 and 2800 MeV. The angular distribution is
smooth and similar in the three cases, with a tail that goes smoothly
to zero for high angles. Increasing the energy just extends the
angular tail of $\Phi(\theta'_1)$ farther and slightly decreases its strength
for low angles, while its maximum is shifted a few degrees to the
right.  Note that the maximum of the angular distribution for these high 
energies is located around 30 degrees.

Thus the origin of the discontinuities observed in Fig.~\ref{fig8bis}
is because the angular distribution $\Phi(\theta'_1)$ has a divergence
or pole for some angle, resulting in a thin peak close to the
pole. When one tries to compute the integral in Eq.~(\ref{angular})
numerically, by evaluating the integrand at some discrete set of
$\theta'_1$ points, sometimes a value close to the pole is reached,
producing the apparent discontinuity.  Trying to integrate the peak
numerically is hard because it is very narrow; so even with many
thousands of points there are still numerical errors.

\begin{figure}
\includegraphics[width=8cm, bb=130 280 410 780]{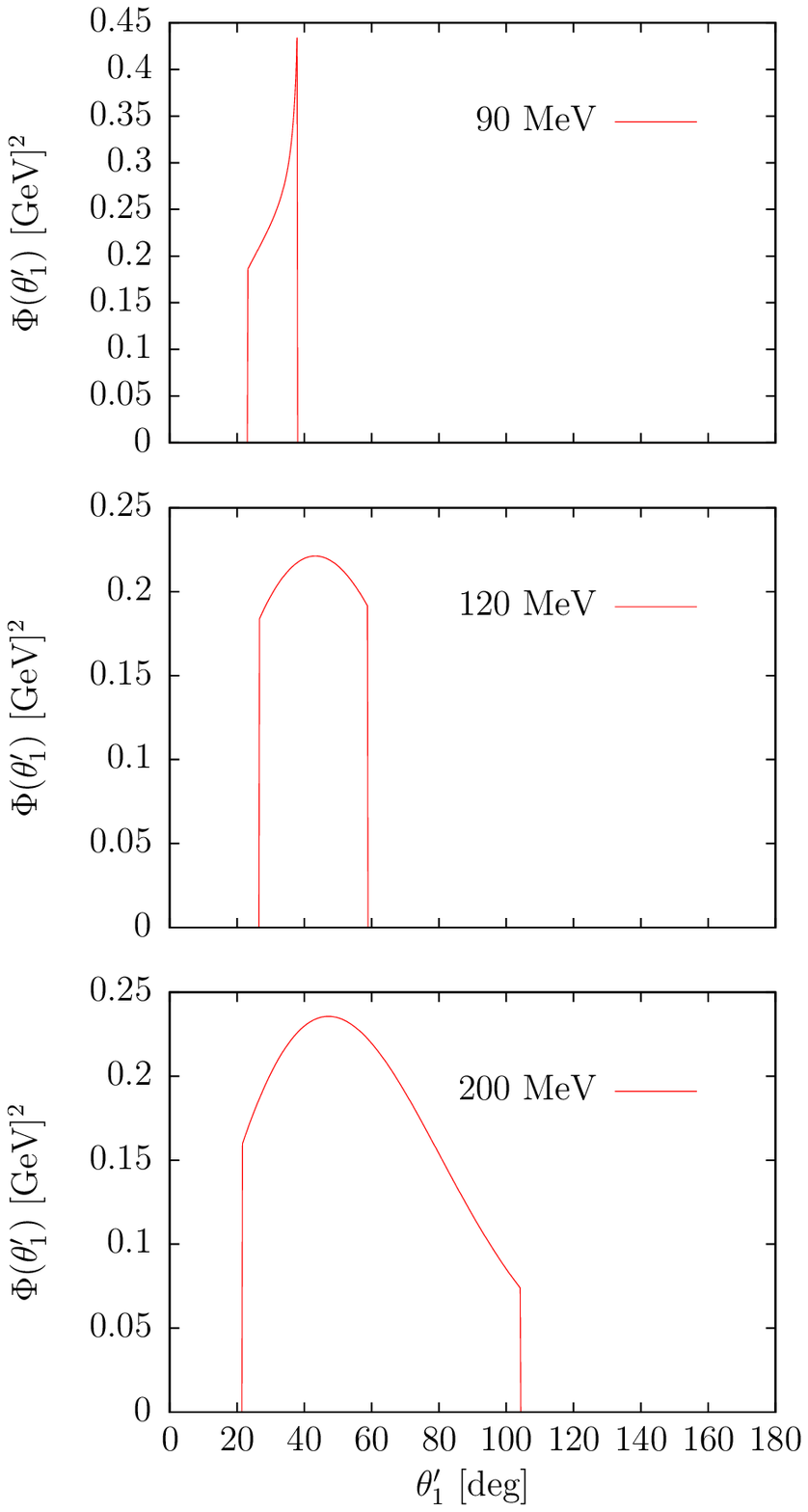}
\caption{
\label{fig13} 
(Color online) 
Angular dependent phase-space function
for $q = 500$ MeV/c, 
for fixed hole momenta  $h_1=h_2=0$, computed for three values of $\omega$
around the quasielastic peak
as a function of the emission angle $\theta'_1$.
}
\end{figure}

\begin{figure}
\includegraphics[width=8cm, bb=130 600 410 780]{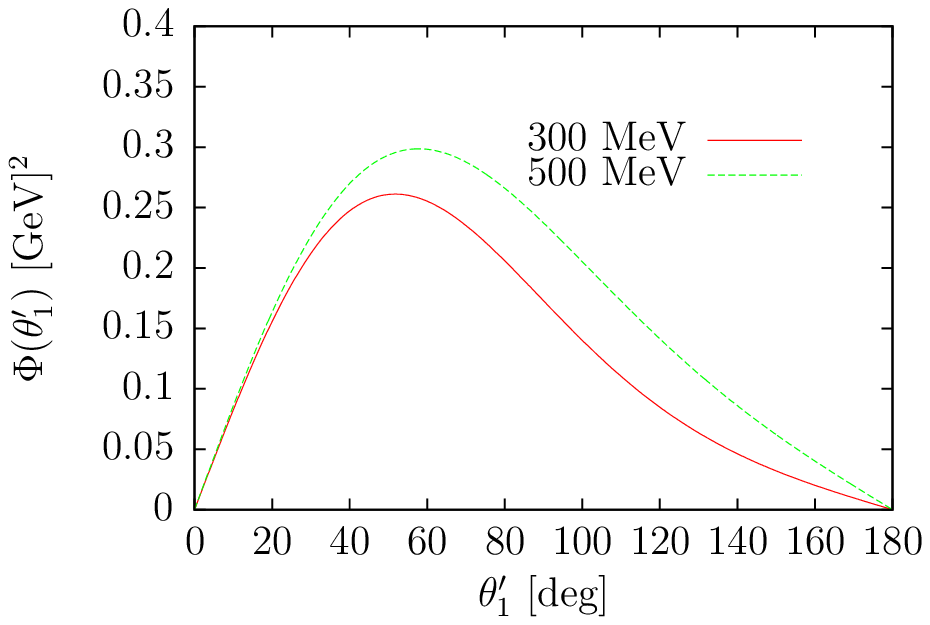}
\caption{
\label{fig14} 
(Color online) 
Angular dependent phase-space function
for $q = 500$ MeV/c, 
for fixed hole momenta  $h_1=h_2=0$, computed for  $\omega=300$ and 500 MeV
as a function of the emission angle $\theta'_1$.
}
\end{figure}

Up to now we have analyzed the problem of the singularity of the
angular distribution for high momentum. Now the question that arises
is why this problem did not apparently emerge when we discussed the
non-relativistic case, that is, for low momentum transfer. The real
fact is that this singularity also appears for low $q$, but only for
very low energy transfer (due to kinematical reasons).  We can see
this in the lower panel of Fig.~\ref{fig8bis}, where we display the
function $\overline{F}(q,\omega)$ in the frozen nucleon approximation
for $q=500$ MeV/c. As before we use 100 integration points.  There is
a narrow peak at threshold followed by rapid, small oscillations.  In
Fig.~\ref{fig13} we show the corresponding angular distribution for
several values of $\omega$. For $\omega=90$ MeV we again see a peak
corresponding to a singularity at the endpoint, but the peak is not as
narrow as for high $q$. Therefore, it can be integrated with few
points. Only for very small $\omega\sim 66$ MeV (not shown in the
figure), we find a very narrow peak.  For higher values of $\omega$
there is no singularity and the angular distribution is smooth and
wide enough to obtain reasonable results with few integration
points. At the QE peak, $\omega\sim 120$ MeV, the angular distribution
is zero outside the interval $25^{\circ} < \theta'_1 < 60^{\circ}$ due
to Pauli blocking, that is also present for $\omega=200$ MeV. For
larger values of $\omega$ (see Fig.~\ref{fig14}), there is no Pauli
blocking and $\Phi(\theta'_1)$ is a smooth distribution with a maximum
that slightly increases with $\omega$ and shifts towards higher
angles.

\subsection{Kinematical analysis}

\begin{figure}
\includegraphics[width=8cm, bb=115 420 395 780]{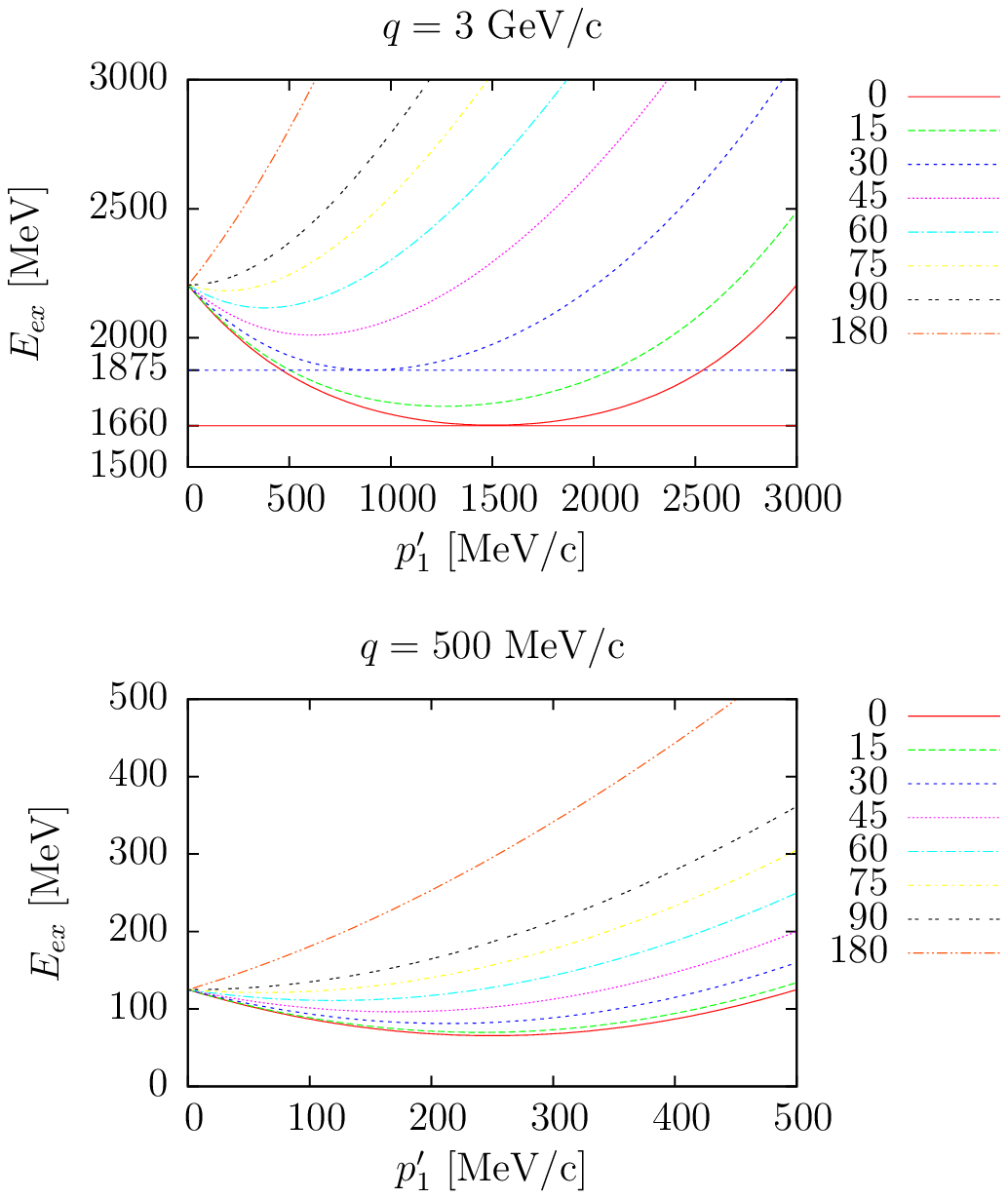}
\caption{
\label{fig15} 
(Color online) 
Plot of the excitation energy of a pair of nucleons at rest
for two values of the momentum transfer and for several emission angles,
as a function of the emission momentum $p'_1$.
}
\end{figure}

We have seen that the angular distribution presents singularities for
some emission angles. The occurrence of the singularity is a consequence of
the kinematical dependence of the excitation energy of the 2p-2h states
\begin{equation}
E_{ex} = E'_1+E'_2-E_1-E_2. 
\end{equation}
In the frozen nucleon limit, it is given by
\begin{equation}\label{Eex}
E_{ex}
= 
\sqrt{p'_1{}^2+m_N^2}+
\sqrt{p'_1{}^2+m_N^2+q^2-2p'_1q\cos\theta'_1}
-2m_N\,,
\end{equation}
which depends on the variables $q,p'_1$ and $\theta'_1$. 
In Fig.~\ref{fig15} we show the value of the excitation energy
as a function of the emission momentum,
for large and intermediate  values of $q$. For each $q$ we plot 
curves for several values of 
the emission angle $\theta'_1$ from 0 to 180$^{\circ}$.

In the upper panel the momentum transfer is $q=3$ GeV/c.  For $p'_1=0$
all of the curves collapse to the quasielastic peak energy.  
The lower horizontal straight line corresponds to
energy $\omega=1660$ MeV, and this is the minimum energy for which
two-particle emission is possible by energy-momentum conservation,
that is, there is a solution of the equation $\omega=E_{ex}(p'_1)$
that corresponds to the intersection point between the
straight line and the lower excitation-energy curve for $\theta'_1=0$,
corresponding precisely to the minimum of the curve.  For angles above
$\theta'_1=0$ two-particle emission is not possible with this excitation
energy. This explains why for very low energy the emission is forward.

If we increase the excitation energy to $\omega= 1875$ MeV,
represented by the upper straight line of Fig.~\ref{fig15}, we see
that it crosses all of the curves below $\theta'_1=30^{\circ}$, that is,
emission is possible only for angles in the interval $[0,30^{\circ}]$.
We also see that for each angle in this interval there are two
values of  $p'_1$ with this excitation energy,
corresponding to the two solutions $p'_1{}^{(\pm)}$,  
Eq.~(\ref{segundo-grado}),
of  the energy conservation equation  $\omega= E_{ex}$.

For $30^{\circ}$ both solutions coincide with
 the position of the minimum of $E_{ex}(p'_1)$. 
A singularity of the angular distribution $\Phi(\theta'_1)$ 
is expected at the end angle $\theta'_1=30^{\circ}$,
because the minimum 
of the curve $E_{ex}(p'_1)$ holds precisely at 
the  solution of the energy conservation equation. Thus 
\begin{equation}
\frac{dE_{ex}}{dp'_1}=0.
\end{equation}
Now the angular distribution is proportional to 
\begin{equation}
\int dp'_1 p'_1{}^2 \delta(E_{ex}-\omega),
\end{equation}
which may be computed by changing 
variables $p'_1 \rightarrow E_{ex}(p'_1)$. Therefore,  
it is proportional to the Jacobian
\begin{equation}
dp'_1= \frac{dE_{ex}}{\left|\frac{dE_{ex}}{dp'_1}\right|} \, ,
\end{equation}
that diverges at the minimum because the denominator is zero.
Note from Fig.~\ref{fig15} that this divergence of the angular
distribution occurs for all the $\omega$ values below the QE
peak, but at a different value of the angle.  This angle must be such
that the corresponding excitation energy curve in Fig.~\ref{fig15} has
its minimum at $E_{ex}=\omega$.
  
Above the quasielastic peak energy $\omega_{QE}$ there is no
divergence because, from Fig.~\ref{fig15}, the minimum is always below
$\omega_{QE}$.  We also see that above $90^{\circ}$ there are no
minima, so divergences only occur for angles below $90^{\circ}$.
This can also be seen in Eq.~(\ref{Eex}): for $\cos\theta'_1 <0$ the
excitation energy increases with $p'_1$.

The same conclusions can be drawn for low momentum transfer.  From the
lower panel of Fig.~\ref{fig15} all of the excitation energy curves for
$q=500$ MeV/c have a minimum below $90^{\circ}$. The main difference
with respect to the high $q$ case is that the quasielastic peak occurs
at very low $\omega$ compared with $q$ and that the minimum $p'_1$ for
large angles is located below $k_F$, and does not contribute to the
angular distribution due to Pauli blocking.  Therefore, there will be
singularities only for very low $\omega$ values.

For smaller values of $q \leq 500$ MeV/c
the minima are always below $k_F$ and
there are no singularities in the angular distribution.

Thus the singularity problem appears only for intermediate to high
$q$.  It could seem that the divergence in the angular distribution
could be observed in a coincidence experiment by fixing the emission
angle and energy transfer at the position of a divergence. However,
this cannot be the case because our discussion is valid only in the
frozen nucleon approximation where the initial nucleons are at rest. In a real
system  an integration over initial momenta
is implied, removing the singularity.

\section{Theoretical analysis of the angular distribution}

\subsection{Allowed angular intervals and divergences}

Our next goal is first to find analytically the angle $\theta'_1$ where the
angular distribution diverges as well as the kind of singularity
(we shall see that the singularity is integrable, as it should be by the definition 
of the phase-space function), and second
to  design a method to compute the
angular integral in the vicinity of the singular point.

We start with the formula for the denominator in the angular
distribution, given by the Jacobian, Eq.~(\ref{jacobiano2}). Using
momentum conservation $\np'_2=\np'-\np'_1$ it can be written in the form:
\begin{equation}
d
\equiv
\frac{p'_1}{E'_1}-\frac{\np'_2\cdot\hp'_1}{E'_2}
=
\frac{E'}{E'_1E'_2}\left(p'_1-\tv E'_1\right) \,,
\end{equation}
where $\tv$ is defined in the Appendix, Eq.~(\ref{tv}).
Using energy conservation, written in the equivalent form (see
Eq.~(\ref{energia-alternativa}) in the Appendix), 
$E'_1=\ta +\tv p'_1$, we arrive at
\begin{equation}
d= 
\frac{E'}{E'_1E'_2}
\left( \tb p'_1 -\tv\ta \right) \,,
\end{equation}
where $\tb$ and $\ta$ have been defined in the Appendix,
Eqs. (\ref{tb},\ref{ta}). The quantity in brackets is the
discriminant in the solution of the second-order equation for the
momentum $p'_1$ given in Eq.~(\ref{segundo-grado}).  Therefore we obtain
\begin{equation}
d= \pm \frac{E'}{E'_1E'_2}\sqrt{D} \,,
\end{equation}
where the relativistic discriminant is defined as 
\begin{equation}
D = \ta^2-\tb m_N^2 \,.
\end{equation}
Using $\tb= 1 -\tv^2$, this can be expressed equivalently as
\begin{equation}
D = m_N^2 \left( \tv^2 - \frac{m_N^2-\ta^2}{m_N^2} \right) \,.
\end{equation}
To make explicit the dependence on the emission angle $\theta'_1$,
implicit in the variable $\tv=\np'\cdot\hp'_1/E'$, we note that the
vector $\hp'_1$ has Cartesian coordinates
\begin{equation}
\hp'_1 = (\sin\theta'_1,0,\cos\theta'_1) \,.
\end{equation}
We recall that we are using the reference system 
 where $\nq$ is in the $z$-axis and that we are
taking $\phi'_1=0$.  Therefore $\hp'_1$ is contained in the 
scattering plane, $xz$.

The scalar product appearing in $\tv$ is 
\begin{equation}
\np'\cdot\hp'_1 = p'_x\sin\theta'_1+p'_2\cos\theta'_1 \,.
\end{equation}
We now define the final momentum vector projected over the scattering plane 
\begin{equation}\label{sprima}
\ns'=(p'_x,0,p'_z) = s' (\sin\alpha,0,\cos\alpha) \,.
\end{equation}
This equation defines $\alpha$
 as the angle between $\ns'$ and $\nq$. 
With this definition, the scalar product can be written
\begin{equation}
\np'\cdot\hp'_1 = 
s'\cos(\theta'_1-\alpha) \,.
\end{equation}
Now the discriminant $D$ can be easily written in terms of the vector $\ns'$
as
\begin{equation}
D = \frac{m_N^2s'{}^2}{E'{}^2}\left[\cos^2(\theta'_1-\alpha)-w_0\right] \,,
\end{equation}
where we have defined the non-dimensional variable
\begin{equation}\label{w0}
w_0= \frac{E'{}^2}{s'{}^2}\left(1-\frac{\ta^2}{m_N^2}\right) \,.
\end{equation}
This development  allows one to write the integral over emission angle $\theta'_1$ 
appearing in Eq.~(\ref{integral7Drel}), for fixed $\nh_1,\nh_2$, as
\begin{eqnarray}
I &\equiv& 
\int_0^\pi d\theta'_1 \sin\theta'_1 \frac{p'_1{}^2}{|d|}
\frac{m_N^4}{E_1E_2E'_1E'_2}
\Theta(p'_1,p'_2,h_1,h_2)
\nonumber\\
&=& 
\int_0^\pi d\theta'_1 \sin\theta'_1 
\frac{m_N^4}{E_1E_2E'_1E'_2}
\Theta(p'_1,p'_2,h_1,h_2)
\nonumber\\
&\times&
\frac{\left(\ta\tv\pm\sqrt{D}\right)^2  \theta(D)}%
     {\frac{\tb^2}{E'_1E'_2}m_Ns'\sqrt{\cos^2(\theta'_1-\alpha)-w_0}} \,,
\end{eqnarray}
where $p'_1= p'_1{}^{(\pm)}= (\ta\tv\pm\sqrt{D})/\tb$ is one of the
solutions of energy conservation. A sum over the two solutions will be
performed later.  The explicit step function $\theta(D)$ indicates that
there is only a solution of energy conservation for a positive value of
$D$, or equivalently, for positive values of the function
\begin{equation} \label{g}
g(\theta'_1)\equiv \cos^2(\theta'_1-\alpha)-w_0.
\end{equation}
Thus we have demonstrated that the  integral $I$ 
has the general form
\begin{equation}
I= 
\int_0^\pi d\theta'_1 \frac{f(\theta'_1)}{\sqrt{g(\theta'_1)}}
\theta(g(\theta'_1)) \,,
\end{equation}
where the function $f(\theta)$ in general has no singularities. This
function will contain the hadronic current when computing the response
functions. The denominator, however, could be zero for some kinematics.
We can consider three cases depending on the value of $w_0$:
\begin{itemize}
\item 
If $w_0>1$ there is no solution of the energy conservation equation, 

\item 
If $w_0<0$ there is always solution of the energy conservation equation. 
All of the
angles are allowed and there is no singularity in the angular
distribution.

\item 
If $0\leq w_0\leq 1$ the angular distribution is different from 
zero only in  one or two angular intervals. The angular distribution is 
infinite for $g(\theta'_1)=0$ or $\cos^2(\theta'_1-\alpha)=w_0$.
\end{itemize}

In the last case
 there are two solutions for this equation given implicitly by 
$\cos  (\theta'_1-\alpha)=\pm\sqrt{w_0}$.
Taking the arc-cosine, we  define the two angles
\begin{equation}
\varphi_1 \equiv \cos^{-1}\sqrt{w_0},
\kern 1cm
\varphi_2 \equiv \cos^{-1}(-\sqrt{w_0}),
\end{equation}
such that $0\leq \varphi_1,\varphi_2<\pi$.
The position of the divergence is defined up to a $\pm\pi$ term
\begin{equation}
\theta'_1-\alpha = \varphi_1\pm\pi,\varphi_2\pm\pi \,.
\end{equation}
To determine the exact position of the divergence 
and the intervals of the allowed angular distribution 
we must analyze the eight possible cases displayed in Fig.~\ref{fig21}.
The eight cases are classified according to the values of $\alpha$ and $w_0$. 
They are the following:

\begin{figure}
\includegraphics[width=8cm, bb=90 265 480 795]{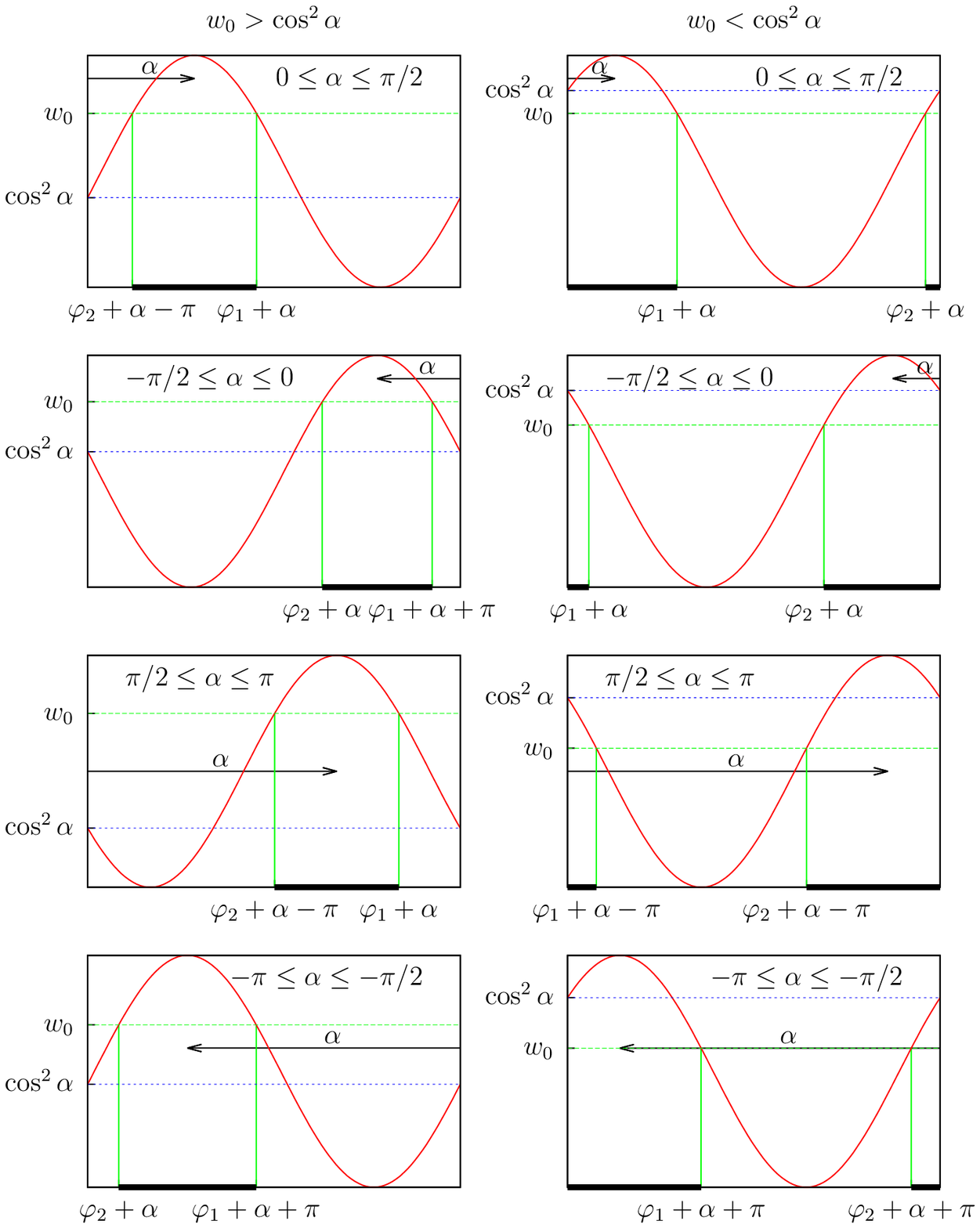}
\caption{
\label{fig21} 
(Color online) 
Plots of the function $\cos^2(\theta'_1-\alpha)$ as a function of
$\theta'_1$, for the geometries of the 8 different cases 
depending on the values of $\alpha$ and
$w_0$. In each panel we show with bold lines the angular intervals
where the integral is performed $\cos^2(\theta'_1-\alpha)> w_0^2$ 
}
\end{figure}

\begin{itemize}

\item
Case (1a): $0\leq\alpha\leq\frac{\pi}{2}$ and $w_0>\cos^2\alpha$.
The angular distribution interval is
\begin{equation} \label{1a}
L \equiv [\varphi_2+\alpha-\pi, \varphi_1+\alpha]
\end{equation} 

\item
Case (1b): $0\leq\alpha\leq\frac{\pi}{2}$ and $w_0<\cos^2\alpha$.
There are two angular distribution intervals 
\begin{equation}
L_1 \equiv [0,\varphi_1+\alpha]\,,
\kern 1cm
L_2 \equiv [\varphi_2+\alpha,\pi] \,.
\end{equation} 

\item
Case (2a): 
$-\frac{\pi}{2}\leq\alpha\leq 0$
and $w_0>\cos^2\alpha$.
The angular distribution interval is
\begin{equation}
L \equiv [\varphi_2+\alpha, \varphi_1+\alpha+\pi] \,.
\end{equation} 

\item
Case (2b): 
$-\frac{\pi}{2}\leq\alpha\leq 0$
and $w_0<\cos^2\alpha$.
There are two angular distribution intervals 
\begin{equation}
L_1 \equiv [0,\varphi_1+\alpha]\,,
\kern 1cm
L_2 \equiv [\varphi_2+\alpha,\pi]\,. 
\end{equation} 

\item
Case (3a): 
$\frac{\pi}{2}\leq\alpha\leq \pi$
and $w_0>\cos^2\alpha$.
The angular distribution interval is
\begin{equation}
L \equiv [\varphi_2+\alpha-\pi, \varphi_1+\alpha] \,.
\end{equation} 

\item
Case (3b): 
$\frac{\pi}{2}\leq\alpha\leq \pi$
and $w_0<\cos^2\alpha$.
There are two angular distribution intervals 
\begin{equation}
L_1 \equiv [0,\varphi_1+\alpha-\pi]\,,
\kern 1cm
L_2 \equiv [\varphi_2+\alpha-\pi,\pi]\,.
\end{equation} 

\item
Case (4a): 
$-\pi \leq\alpha\leq -\frac{\pi}{2}$
and $w_0>\cos^2\alpha$.
The angular distribution interval is
\begin{equation}
L \equiv [\varphi_2+\alpha, \varphi_1+\alpha+\pi] \,.
\end{equation} 

\item
Case (4b): 
$-\pi \leq\alpha\leq -\frac{\pi}{2}$
and $w_0<\cos^2\alpha$.
There are two angular distribution intervals 
\begin{equation}  \label{4b}
L_1 \equiv [0,\varphi_1+\alpha+\pi]\,,
\kern 1cm
L_2 \equiv [\varphi_2+\alpha+\pi,\pi]\,.
\end{equation} 

\end{itemize}

Note that only the cases 1 and 2 are possible for large $q> 2k_F$,
which is the case of most interest for neutrino and electron scattering 
applications at intermediate energies.
Cases 3 and 4 are only possible for low $q$, where the non
relativistic formalism can be applied. They are given here for
completeness.

\subsection{Integration of divergences}

Two singularities appear in the angular distribution at the boundaries
of the allowed intervals, corresponding to
$\cos^2(\theta'_1-\alpha)=w_0$.  

To integrate the resulting function numerically is not simple because
the width of the infinite peak is small around the asymptote  
and a very small step is needed. However the divergence is integrable.
The situation is similar to
performing the integral of the function $1/\sqrt{x}$ between 0 and 
$\epsilon>0$
\begin{equation} \label{epsilon}
\int_0^{\epsilon} \frac{dx}{\sqrt{x}} 
= \left. 2\sqrt{x} \right|_0^{\epsilon} = 2\sqrt{\epsilon} \,.
\end{equation}
The integrand is infinite for $x=0$, but the integral is well defined 
because the function  increases slower than $x^{-1}$. 

In our case we exploit the above property of the integral of
$1/\sqrt{x}$, that is, we perform the integral around the divergence
analytically by assuming that the numerator does not change too much in a
small interval.

Specifically, we consider the case (1a), where the integration
interval $[\theta_1,\theta_2]$ is given in Eq.~(\ref{1a}) and there
are two singularities at the ends of the interval.
We are then involved with an integral of the kind
\begin{eqnarray}
I(\theta_1,\theta_2)
&\equiv&
 \int_{\theta_1}^{\theta_2}
\frac{f(\theta) d\theta }{\sqrt{g(\theta)}}
\\
&=& 
I(\theta_1,\theta_1+\epsilon)+
I(\theta_1+\epsilon,\theta_2-\epsilon)+
I(\theta_2-\epsilon,\theta_2) \,.
\nonumber
\end{eqnarray}
We have written this equation as the sum of three integrals.
Here $\epsilon$ is a small number that will allow us to integrate analytically 
around the divergence points by exploiting Eq.~(\ref{epsilon}).
First we  re-write the integrand by multiplying and dividing by the derivative
$g'(\theta)=dg/d\theta$, as
\begin{equation}
\frac{f(\theta)}{\sqrt{g(\theta)}}
=
2 \frac{f(\theta)}{g'(\theta)}
\frac{d \sqrt{g(\theta)}}{d\theta} \,.
\end{equation}
Under the assumption that the function
$\frac{f(\theta)}{g'(\theta)}$ is finite and almost constant in the 
small interval $[\theta_1,\theta_1+\epsilon]$,
the integral around the first singular point
can be approximated by
\begin{eqnarray}
I(\theta_1,\theta_1+\epsilon)
&=&
\int_{\theta_1}^{\theta_1+\epsilon}
\frac{f(\theta) d\theta }{\sqrt{g(\theta)}}
\nonumber\\
&\simeq&
2 \frac{f(\theta_1)}{g'(\theta_1)}
\int_{\theta_1}^{\theta_1+\epsilon}
\frac{d \sqrt{g(\theta)}}{d\theta}
 d\theta 
\nonumber\\
&=&
2 \frac{f(\theta_1)}{g'(\theta_1)} \sqrt{g(\theta_1+\epsilon)}
\end{eqnarray}
because $g(\theta_1)=0$. 
This is a result that already can be used in practice to 
compute the integral around the divergence. However, we prefer 
to write it in an equivalent way that is valid for the eight cases.
Using the fact that $\epsilon$ is small, we first expand 
$g(\theta_1-\epsilon)\simeq -g'(\theta_1)\epsilon$. 
Therefore
\begin{equation}
I(\theta_1,\theta_1+\epsilon)
 = \frac{2f(\theta_1)}{\sqrt{g'(\theta_1)}}\sqrt{\epsilon} \,.
\end{equation}
 From the definition of $g(\theta)$, Eq.~(\ref{g}), 
\begin{equation}
g'(\theta)=-2\cos(\theta-\alpha)\sin(\theta-\alpha)
\end{equation}
using $\theta_1=\varphi_2+\alpha-\pi$, and
we get the following  values at the divergence:
$\cos(\theta_1-\alpha)=\sqrt{w_0}$,
and $\sin(\theta_1-\alpha)=-\sqrt{1-w_0}$. We obtain for the derivative at
$\theta_1$
\begin{equation}
g'(\theta_1)= 2\sqrt{w_0(1-w_0)}\,.
\end{equation}
The integral around the singular point $\theta_1$ 
can be finally written as
\begin{equation} \label{singularity}
I(\theta_1,\theta_1+\epsilon) =
\frac{ f(\theta_1)\sqrt{2\epsilon}}{[w_0(1-w_0)]^{1/4}} \,.
\end{equation}
A similar calculation gives for the integral around the upper
divergence angle $\theta_2$ the result:
\begin{equation}
I(\theta_2-\epsilon,\theta_2) \simeq
-\frac{2f(\theta_2)}{g'(\theta_2)}\sqrt{g(\theta_2-\epsilon)}
\simeq
\frac{f(\theta_2)\sqrt{2\epsilon}}{[w_0(1-w_0)]^{1/4}} \,.
\end{equation}
Finally, we can write the integral as
\begin{equation}
I(\theta_1,\theta_2)
= I(\theta_1+\epsilon,\theta_2-\epsilon)+
\frac{[f(\theta_1)+f(\theta_2)]\sqrt{2\epsilon}}{[w_0(1-w_0)]^{1/4}} \,.
\end{equation}
The integral $I(\theta_1+\epsilon,\theta_2-\epsilon)$ can now be 
evaluated numerically.

A systematic analysis of the eight cases (a1)--(4b) 
shows that this result can be extended
for all kinematics. That is, the contribution from the neighborhood of
a singularity $\theta_1$ is given by Eq.~(\ref{singularity}), where
$\epsilon$ is a small integration interval to the left or to the right of
the divergence point.

\section{Results for the 2p-2h phase-space function}

Here we present results for $F(q,\omega)$ using the integration method
introduced in the previous section. It results in the following
integration algorithm: For each pair of holes $\nh_1,\nh_2$ we first
compute the variable $w_0$, Eq.~(\ref{w0}).  According to the previous
section, if $w_0>1$, there is no solution of the energy conservation
equation and consequently, this pair of holes does not contribute to
$F(q,\omega)$. If $w_0<0$, all of the emission angles are allowed for
the first particle, so we can safely compute the integral over
$\theta'_1$ numerically in the interval $[0,\pi]$. If $0\leq w_0\leq 1$ then we
compute the angle $\alpha$ defined in Eq.~(\ref{sprima}) and determine
the case (1a)--(4b) to which these kinematics belong, and the
corresponding allowed intervals, Eqs. (\ref{1a}--\ref{4b}). We
integrate numerically within each one of the allowed intervals, up to
a distance $\epsilon$ to the singular point.  The integral around the
singular point is made using the semi-analytical method discussed in
the previous section.  Each singular point contributes with a term
given by Eq.~(\ref{singularity}) which we add to the numerical
integral.  We use the value $\epsilon=0.01$, but  
we have checked that the results do not depend on $\epsilon$. 
For the numerical integrals we use Simpson method.

\begin{figure}
\includegraphics[width=8cm, bb=130 280 410 780]{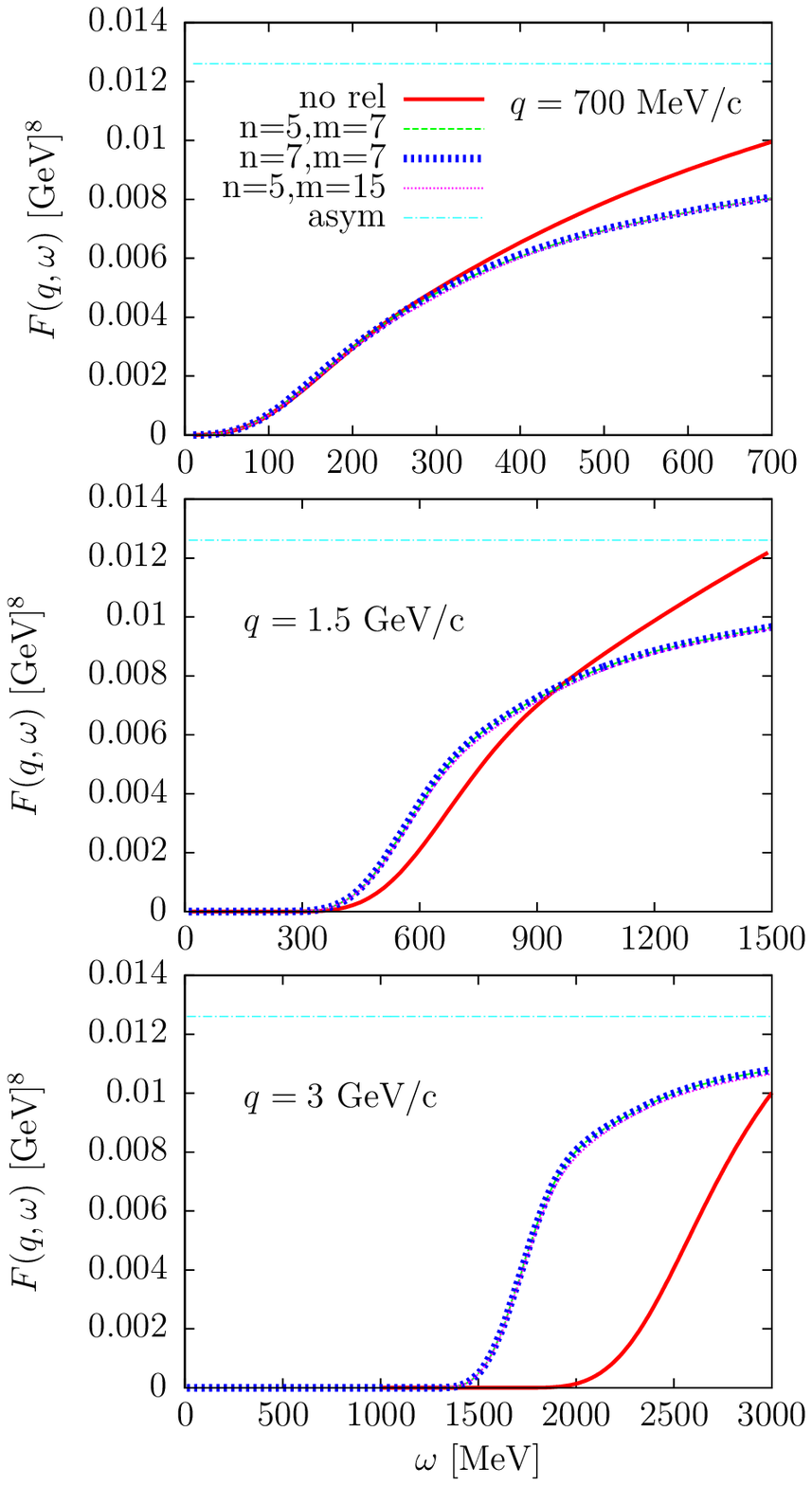}
\caption{
\label{fig24} 
(Color online) 
Total phase-space function for three values of the momentum transfer.  The
number of integration points in each dimension in the hole variables
is indicated by $n$. The number of integration points over the
emission angle $\theta'_1$ is indicated as $m$.  We also show the 
non-relativistic, exact result and the relativistic asymptotic value.  }
\end{figure}

\begin{figure}
\includegraphics[width=8cm, bb=130 280 410 780]{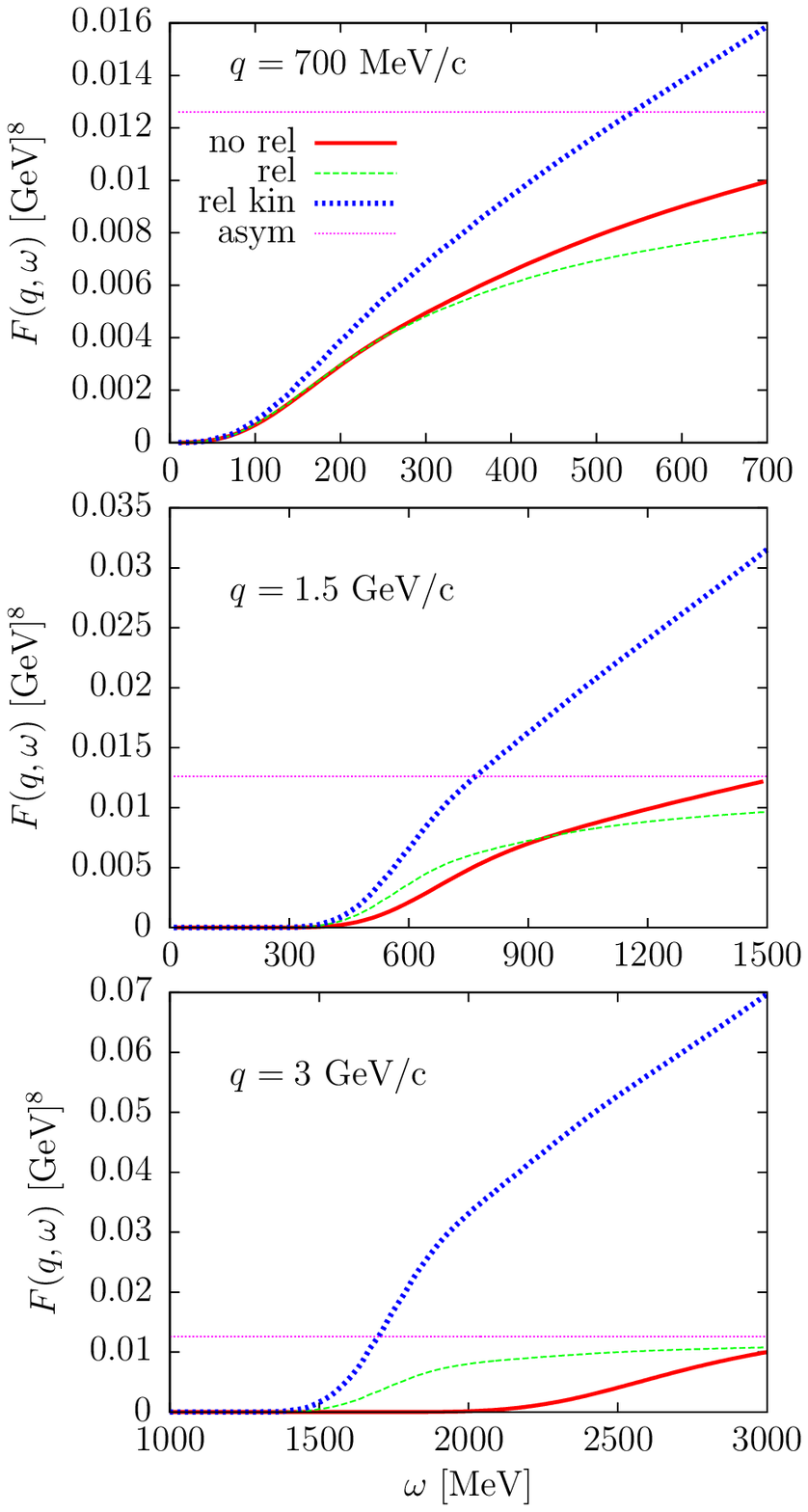}
\caption{
\label{fig25} 
(Color online) 
Effect of implementing relativistic kinematics in a non-relativistic 
calculation of $F(q,\omega)$. Solid lines: non-relativistic result.
Thick dotted lines: relativistic kinematics only  without the 
relativistic factors $m_N/E$. Thin dashed lines: fully relativistic result.  
}
\end{figure}

\begin{figure}
\includegraphics[width=8cm, bb=130 280 410 780]{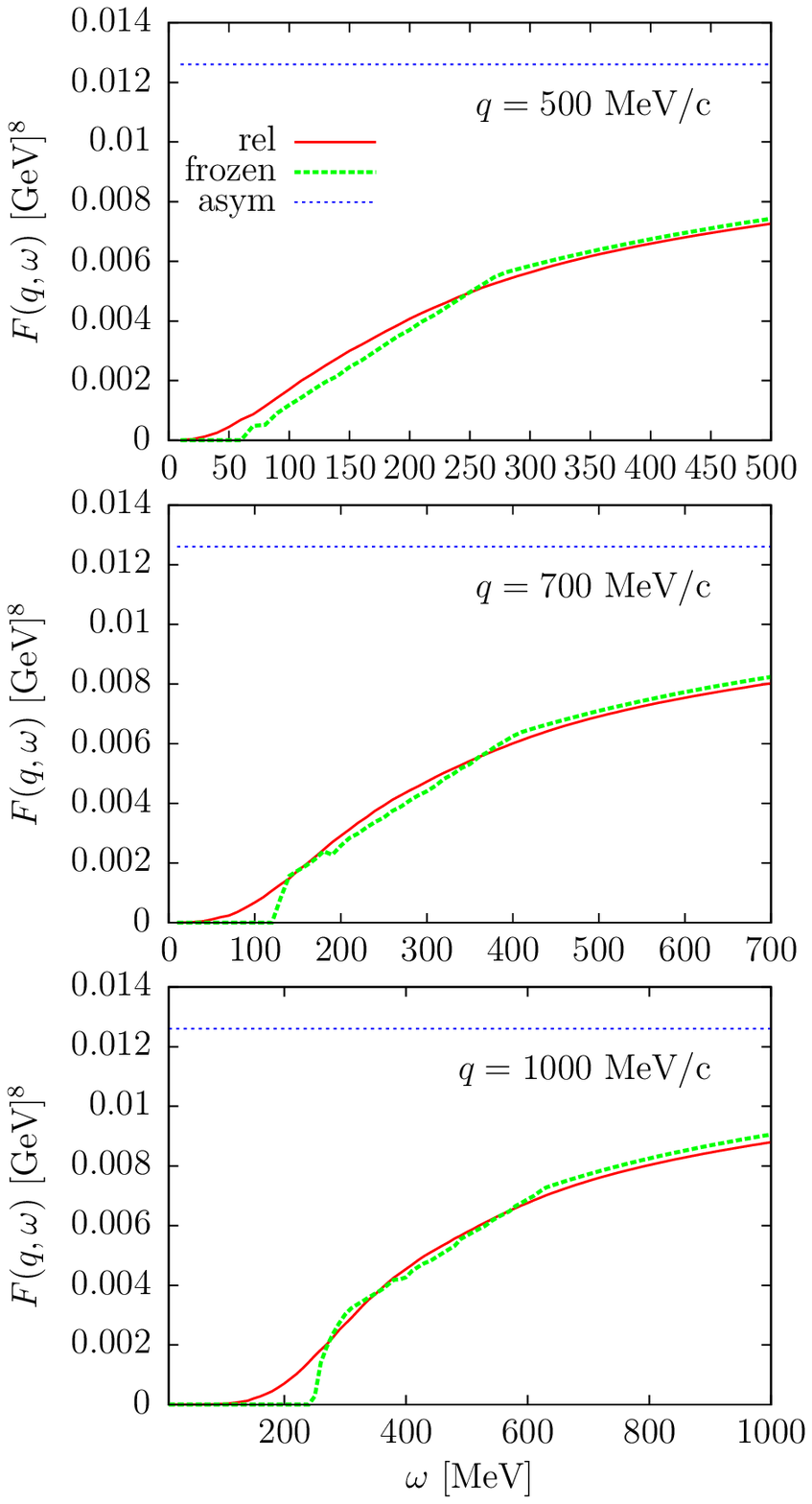}
\caption{
\label{fig26} 
(Color online) 
Relativistic phase-space function $F(q,\omega)$ compared with the frozen
nucleon approximation for low to intermediate momentum transfer.  }
\end{figure}

\begin{figure}
\includegraphics[width=8cm, bb=130 280 410 780]{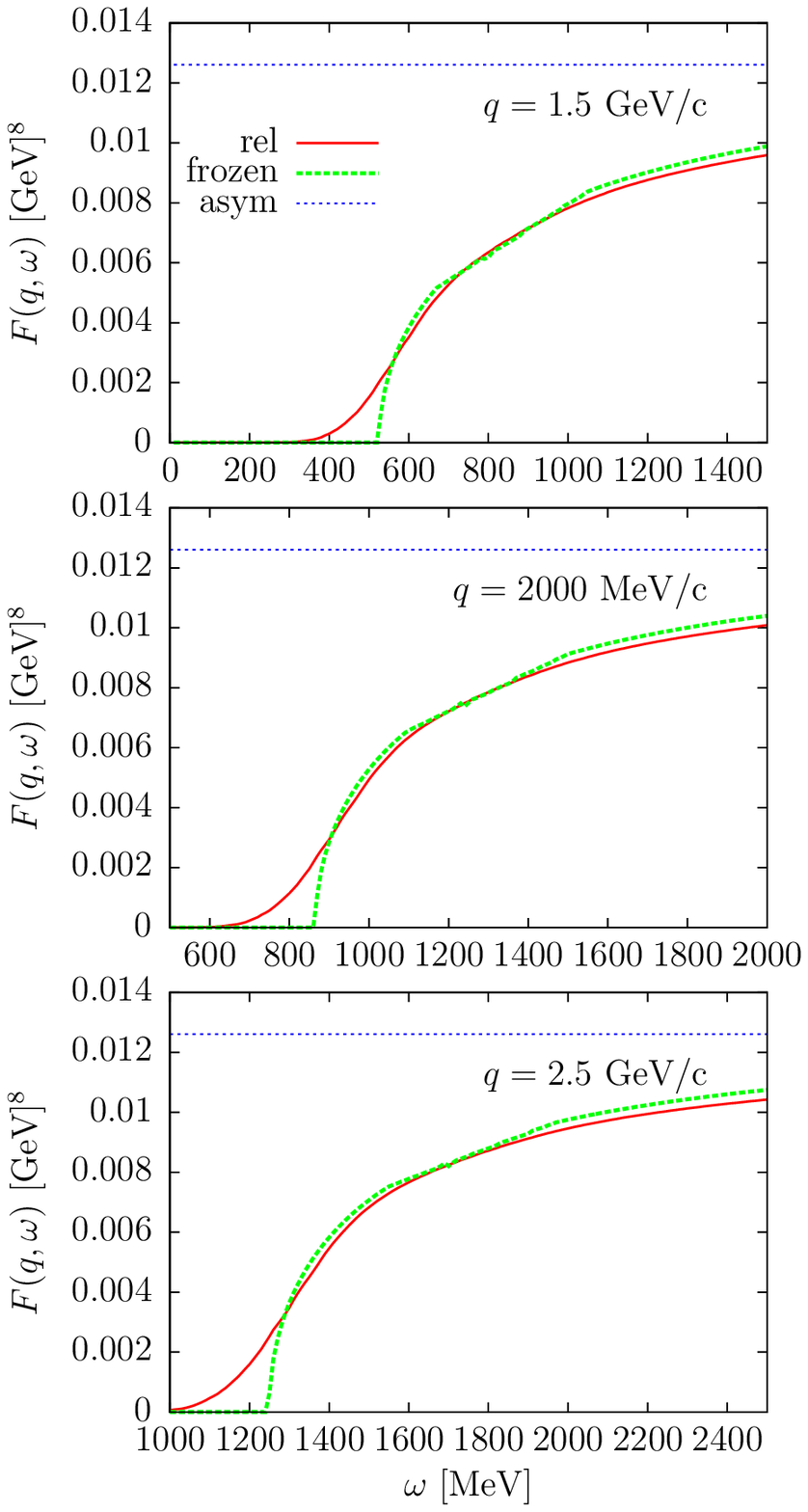}
\caption{
\label{fig27} 
(Color online) 
Relativistic phase-space function $F(q,\omega)$ compared with the frozen
nucleon approximation for high momentum transfer.
}
\end{figure}

In Fig.~\ref{fig24} we show the total phase-space function
$F(q,\omega)$ for three values of the momentum transfer, $q=700$, 1500
and 3000 MeV/c. We study the convergence of the 7D integral. For the
integral over the two holes $\nh_1,\nh_2$, we show results for $n=5$
and 7 points for each dimension. For the inner integral over the
emission angle we use $m=7$ and 15 points. We see that using
$(n,m)=(5,7)$ there is almost no difference with the other cases
$(7,7)$ and $(5,15)$.  As we have seen, the new algorithm allows us to compute 
with small error the inner integral over
$\theta'_1$ using only 7 points. The
dependence on the hole momenta, $\nh_1,\nh_2$, of the resulting
function is very smooth and can be safely computed with a small number
of integration points. The fact that very precise results can be obtained using
$(n,m)=(5,7)$ is an important improvement over previous approaches,
taking into account that the total number of points is $5^6\times 7
\simeq 10^5$, that is, two orders of magnitude less than $10^7$ 
(10 points for each dimension). Thus the computational time 
when we include the nuclear current matrix
elements  will be considerably reduced.

In Fig.~\ref{fig24} we also show the non-relativistic, exact result,
computed using the semi-analytical expression, Eq.~(\ref{fasico2D}).
For $q=700$ MeV/c both relativistic and non-relativistic results coincide
for low energy $\omega<300$ MeV. Above this energy the relativistic
result is below the non-relativistic one.  For $q=1.5$ GeV/c there are
clear differences between the two results for all energies. For high
momentum transfer $q=3$ GeV/c, they are completely different. The 
non-relativistic function is pushed toward higher energies due to the
quadratic momentum dependence of the non-relativistic kinetic
energy. Thus for $q=1.5$ GeV/c, the relativistic results are above
(below) the non-relativistic ones for low (high) energy. For $q=3$
GeV/c the relativistic results are above for
all $\omega$ values allowed.

In all cases $F(q,\omega)$ is below the asymptotic value, 
Eq.~(\ref{asymptotic2}), also shown in Fig.~\ref{fig24}.

In Fig.~\ref{fig25} we get a deeper insight into the size of
relativistic effects. There we show results for $F(q,\omega)$ computed
using relativistic kinematics only, but without including the
relativistic Lorentz-contraction factors $m_N/E$ in particles and
holes. The results increase a lot with respect to the non-relativistic
ones. This is related to the fact pointed out after
Eq.~(\ref{asymptotic2}) for the asymptotic limit of the relativistic
phase-space integral. Without the Lorentz factors, the function
$F(q,\omega)$ would increase as $\omega^2$. This seems to indicate
that in order to ``relativize'' a non-relativistic 2p-2h model,
implementing only relativistic kinematics is not sufficient, since it
goes in the wrong direction.  In fact, results in Fig.~\ref{fig25}
show that the effects coming solely from the relativistic kinematics
lead to differences even larger than the discrepancy between the
non-relativistic and the fully relativistic calculations. 
Therefore, it is essential also to include the Lorentz factors
$m_N/E$.

Note that the behaviour of relativistic effects in the 1p-1h channel
goes in the opposite direction to the one discussed here in the 2p-2h
channel. In fact, in \cite{Ama02} it was shown that implementing
relativistic kinematics without the $m_N/E$ factors in the non
relativistic 1p-1h response function, gives a result closer to the
exact relativistic response function (see Fig. 33 of \cite{Ama02}).

In Figs.~\ref{fig26} and \ref{fig27} 
 we present the results of a study of the validity of the frozen nucleon
approximation to compute $F(q,\omega)$ in a range of momentum transfers. 
This approximation was
introduced for high momentum transfer $q=3$ GeV/c, neglecting the
momenta of the two holes inside the 7D integral, thus reducing it to a
1D integral over the emission angle $\theta'_1$. In Fig.~\ref{fig27} 
the momentum transfer is still high and the frozen nucleon approximation
remains valid. In Fig.~\ref{fig26}
the values of $q$ are not so large, and one could think that the
frozen nucleon approximation is not valid. However, the results of
Fig.~\ref{fig26} demonstrate that it is still a good approximation for
moderate momentum transfer except for very low energy transfer, where
the function $F(q,\omega)$ is small. This is a promising result: 
if the frozen nucleon approximation could be extended to the full response functions when including the
nuclear current, this would mean that the
2p-2h cross section could be approximated by 1D integrals over the
emission angle which would be easy and fast to compute.  In particular,
calculations of this kind could be implemented in existing Monte-Carlo codes.

\begin{figure}
\includegraphics[width=8cm, bb=130 280 410 780]{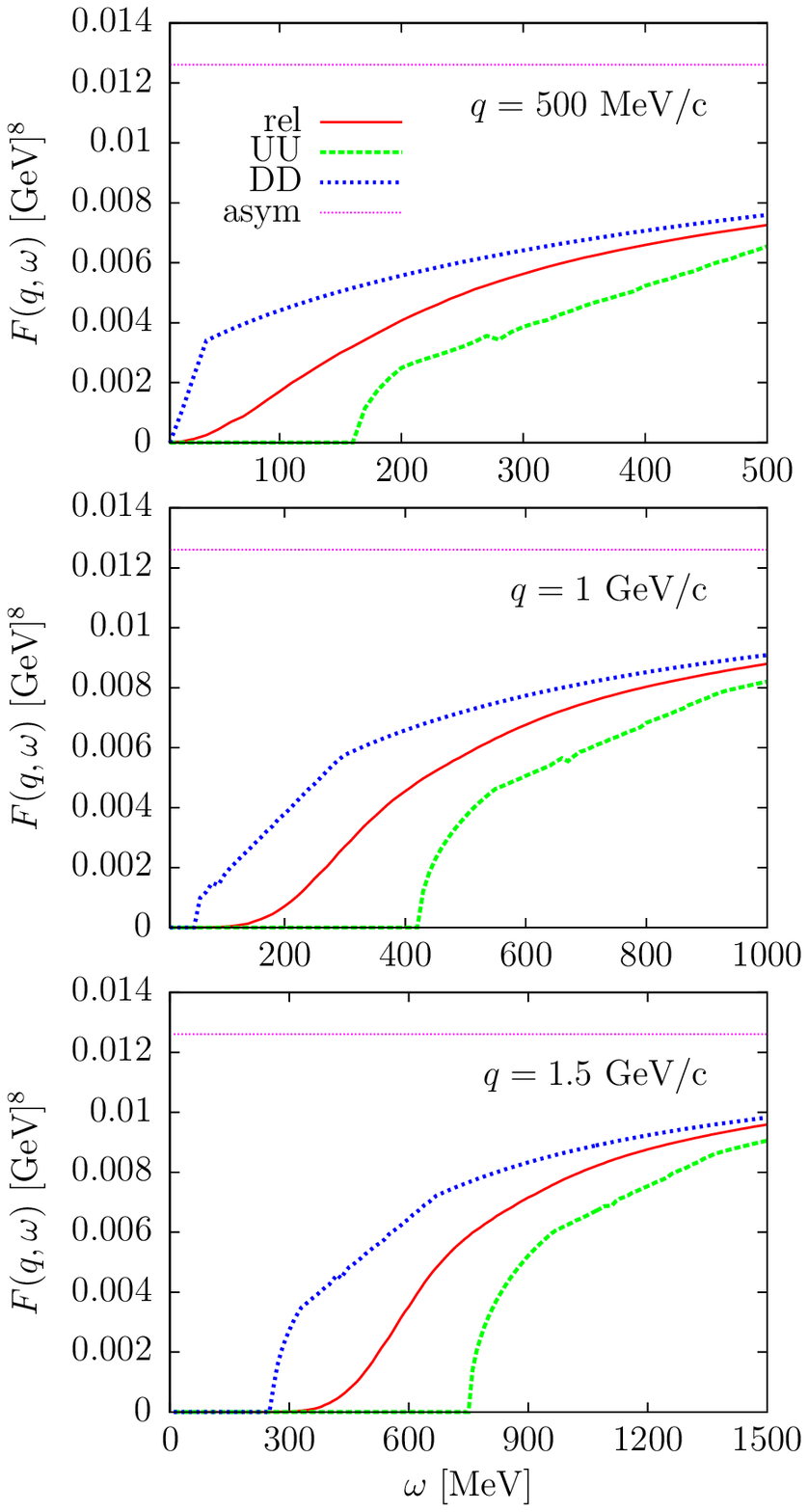}
\caption{
\label{fig28} 
(Color online) 
Relativistic phase-space function $F(q,\omega)$ compared with the 
average-momentum approximation $\overline{F}(q,\omega)$ for a pair of nucleons 
with momentum $200$ MeV/c. In the UU configuration,  both nucleons 
move  along $\nq$  (up). In the DD configuration, both move 
 opposite to $\nq$ (down).
}
\end{figure}

\begin{figure}
\includegraphics[width=8cm, bb=130 280 410 780]{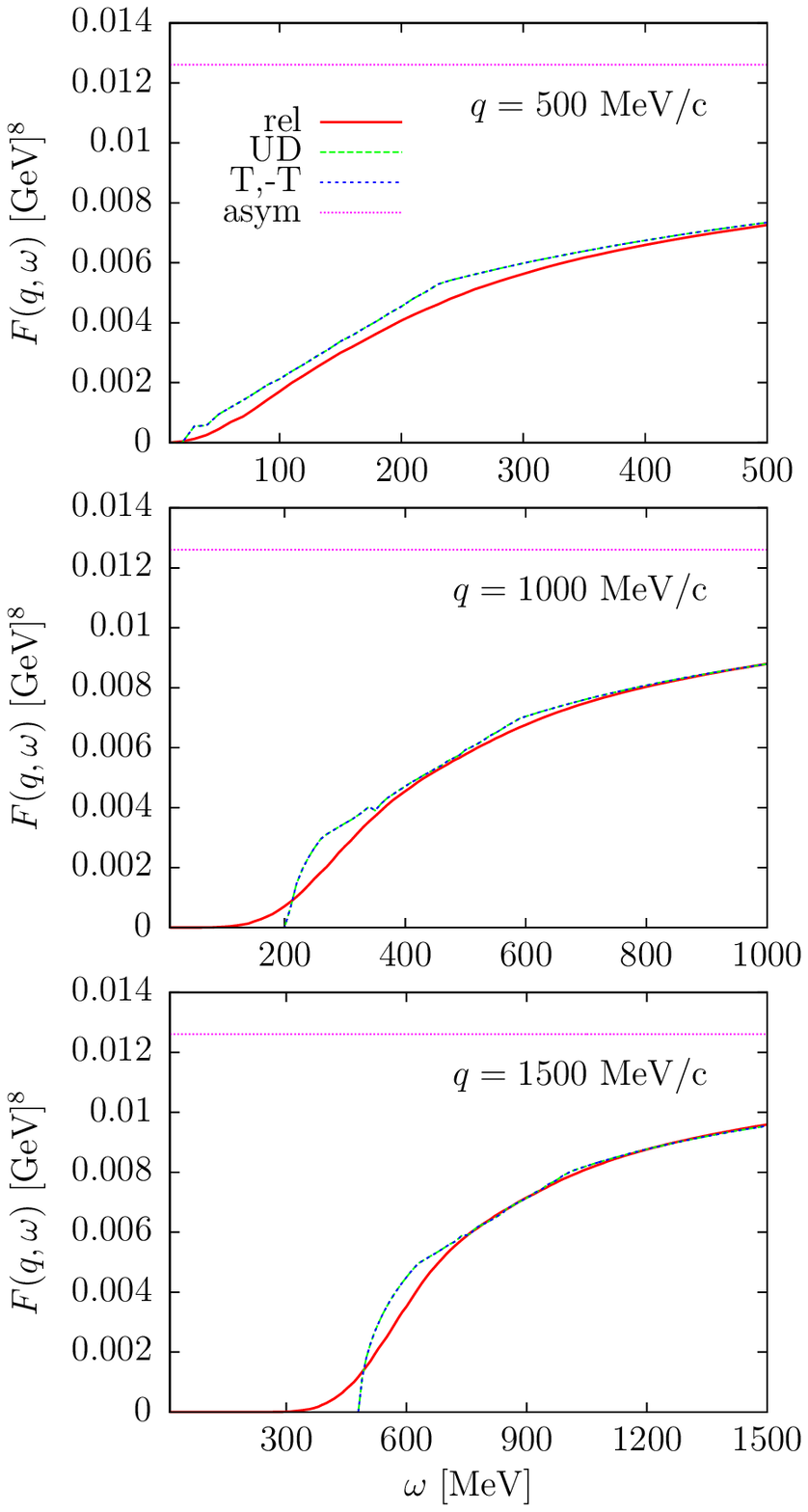}
\caption{
\label{fig29} 
(Color online) Relativistic phase-space function $F(q,\omega)$
compared with the average-momentum approximation
$\overline{F}(q,\omega)$ for a pair of nucleons with momentum $200$
MeV/c pointing in opposite directions (total momentum equal to zero).
In the UD configuration one moves along $\nq$ (U) and the other
opposite to $\nq$ (D). In the T,-T configuration one moves in the
$x$-direction and the other in the $-x$-direction.  }
\end{figure}

\begin{figure}
\includegraphics[width=8cm, bb=130 280 410 780]{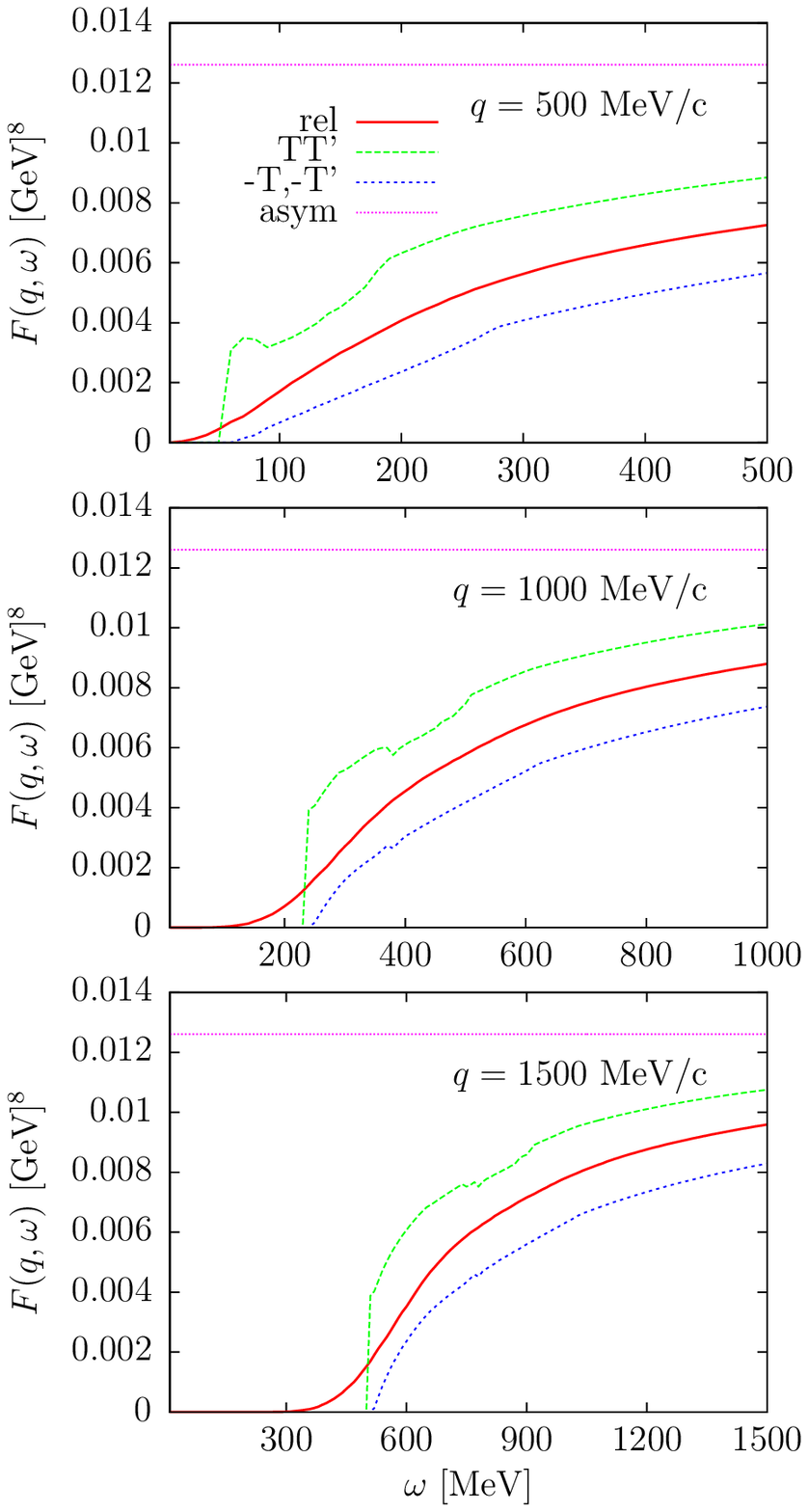}
\caption{
\label{fig30} 
(Color online) 
Relativistic phase-space function $F(q,\omega)$ compared with the 
average-momentum
approximation $\overline{F}(q,\omega)$ for a pair of nucleons with
momentum $200$ MeV/c pointing in perpendicular directions.
 In the T,T' configuration one moves along $x$ (T) and
the other along to $y$ (T'). In the -T,-T' configuration
they move along the $-x$ and $-y$ directions, respectively.
}
\end{figure}

\begin{figure}
\includegraphics[width=8cm, bb=85 240 525 760]{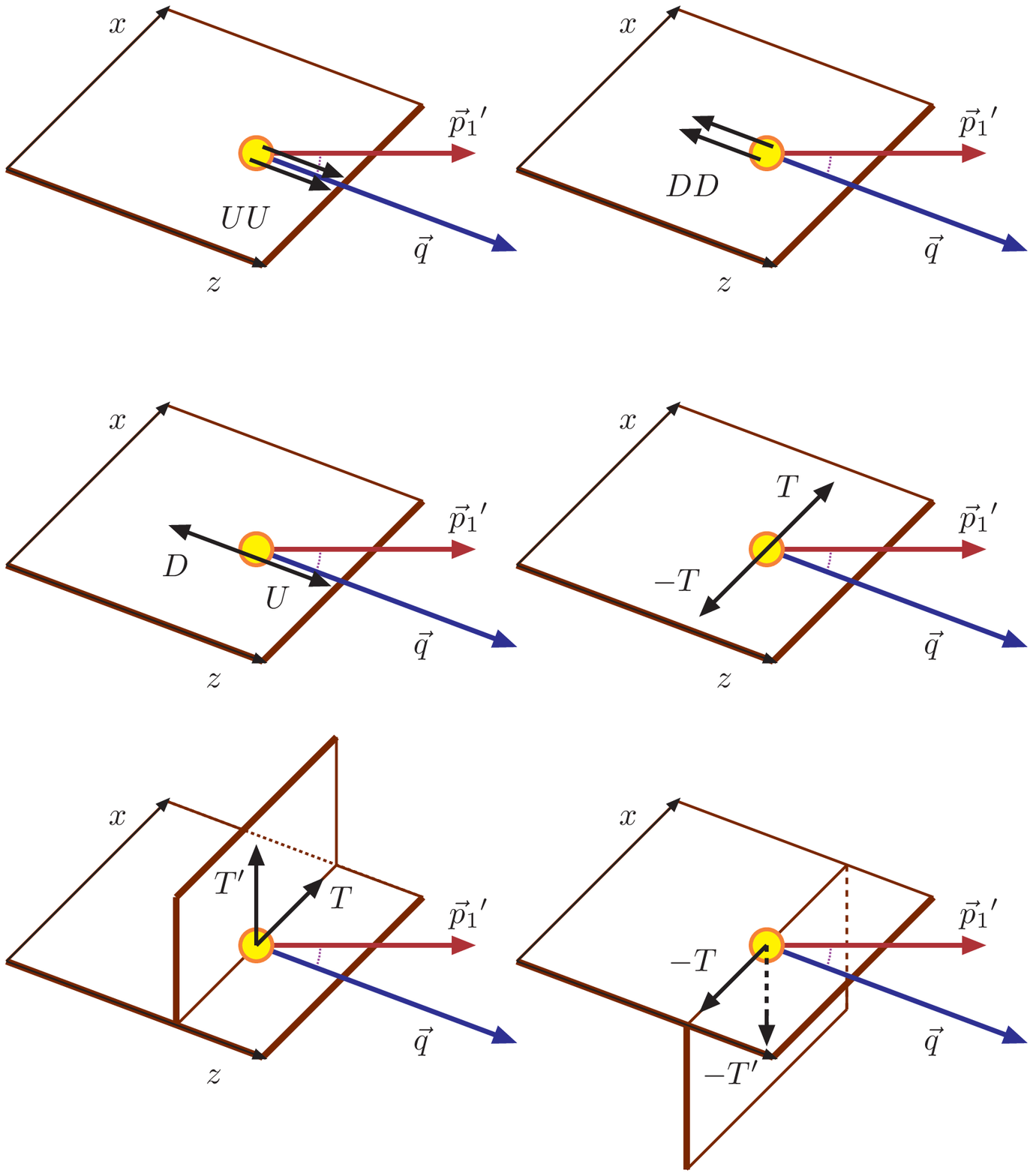}
\caption{
\label{fig31} 
(Color online) 
Geometry employed
for emission of a pair of nucleons with momenta parallel  (cases $UU$, $DD$), 
anti-parallel (cases $UD$, $T-T$) and perpendicular (cases $TT'$, $-T-T'$).
}
\end{figure}

To illustrate the reasons why the frozen nucleon approximation works
for moderate momentum transfer we present
Figs.~\ref{fig28}, \ref{fig29} and \ref{fig30}. We compare
$F(q,\omega)$ with the barred phase-space function
$\overline{F}(q,\omega)$, defined in Eq.~(\ref{barred}), computed for
several $(\nh_1,\nh_2)$ configurations.  The 
{ \sl ``average-momentum approximation''} 
is
similar to the frozen nucleon approximation in the sense that 
the two hole momenta $\nh_1$, $\nh_2$ are set to a constant inside
the integral. For a pair configuration $(\nh_1,\nh_2)$, the function
$\overline{F}(q,\omega)$ gives the contribution of such a pair to the
phase-space function, multiplied by $V_F^2$, where $V_F=4\pi k_F^3/3$ is
the volume of the Fermi sphere. The total $F(q,\omega)$ is the sum of
the contributions from all of the pairs, or equivalently the average of
all of the barred functions $\overline{F}(q,\omega)$ over the different pair
configurations. 

In Fig.~\ref{fig31} we show the geometry for the configurations used
in Figs.~\ref{fig28}--\ref{fig30}. For low values of the momenta
$h_1,h_2$, the frozen nucleon approximation should be a good approximation to
the average phase-space. For larger values of the momenta, we find
pairs of configurations with opposite total momentum $\np=\nh_1+\nh_2$
that contribute above or below the average in approximately equal
footing, so they do not change the mean value very much. 

In the  first example, Fig.~\ref{fig28},  we show the 
contribution of two pairs of
nucleons with the same momentum $h_1=h_2=200$ MeV/c, and both
parallel, pointing upwards ($U$) and downwards ($D$) with respect to
the $z$-axis, that is, the direction of
$\nq$. The contribution of the $UU$ configuration is smaller than
average, while the $DD$ is larger.  This is so because
in the $UU$ case, the total momentum $p'$ in the final state is
large. By momentum conservation, the momenta $p'_1$ and $p'_2$ must also
be large. Therefore these states need a large excitation
energy, and they start to contribute for high $\omega$ transfer.  In
the DD configuration the total momentum $p'$ is small, so the final
momenta $p'_1$ and $p'_2$ can also be small, will small excitation
energy. Therefore they start to contribute at very low $\omega$.

In the example of Fig.~\ref{fig29} two anti-parallel configurations are
shown. In the $UD$ case one nucleon is moving upwards and the other
downwards the $z$-axis with total momentum zero of the pair. This
situation is similar to that of a pair of highly correlated nucleons
with large relative momentum~\cite{Kor14}. 
Since the total momentum is zero, the
final 2p-2h state has total momentum $\nq$, exactly the same that it
would have in the frozen nucleon approximation. Therefore the contribution of
this configuration is similar to the average. The same conclusions can
be drawn in the case of the configuration $T,-T$, with one nucleon
moving along the $x$-axis (transverse direction) and the other along
$-x$ with opposite momentum. The contribution of this pair is exactly
the same as that of the $UD$ configuration in the total phase-space function.

Finally we show in Fig.~\ref{fig30} two intermediate cases that are  neither
parallel nor anti-parallel configurations. They consist in two pairs of
transverse nucleons moving along mutually perpendicular directions.  
In the
first case we consider a $T$-nucleon and a second $T'$-nucleon moving in the
$y$-axis out of the scattering plane.  The contribution of the $TT'$
pair is large, while the one of the opposite case, $-T,-T'$ is small. On
the average they are close to the total result.

\section{Perspectives on the calculation of 
2p-2h electroweak response functions}

The next step in our project of an exact evaluation of the
relativistic 2p-2h electroweak response functions in the Fermi gas
model initiated with the approach in the present paper would be to apply it to a
more realistic situation, {\it i.e.}, electron and neutrino
scattering.  The 2p-2h states can be excited by two-body MEC
operators, involving exchange of an intermediate meson between two
nucleons.  A complete calculation, including all of the MEC diagrams
with one pion exchange, is out of the scope of the present paper, and
will be reported in a forthcoming publication.  However in this
section we discuss the perspectives opened by the formalism
presented here.

One question can arise on why the integration problem related to the
divergence in the angular distribution, that is the central issue in
this work does not appear in other approaches. The models developed by 
Martini~\cite{Mar09} and Nieves~\cite{Nie11} neglect the
direct-exchange interference terms in the hadronic tensor.  In this
approximation in the non-relativistic case, the change of variables
introduced in Eqs. (\ref{variables1},\ref{variables2}) reduces the
integration to 2D, the integration variables in this case are
proportional to the magnitude of the transferred momenta to the two
nucleons. In the relativistic model of \cite{Nie11} an additional
approximation is made for the total $WNN$ interaction vertex, where
the dependence on the initial nucleon momentum is neglected, by fixing
it to an average value over the Fermi sea. This trick allows one to
factorize the two Lindhard functions linked to the two nucleon loops
in the many-body diagrams. Thus two approximations are required in
this case to reduce the calculation to a 4D integral over the
four-momentum of one of the exchanged pions.  The exact calculation
including all the terms in the hadronic tensor (direct, exchange
and interference) requires the complete 7D integral. 

Obviously a change of variables can be made to elliminate the
divergence.  One possibility is to make the change
$\theta'_1\rightarrow \sqrt{g(\theta'_1)}$, where $g(\theta'_1)$ is
the function defined in Eq. (\ref{g}). This corresponds to the change of
variables made in Sect. VIB to integrate analitically around the
divergence.

The standard way to handle this problem in the Monte Carlo generators
\cite{Sob12} is to compute the 2p phase-space angular distribution in
the center of mass (CM) system of the final nucleons, because it is
angular independent, although Pauli blocking can forbid some
angular regions. A transformation to the Lab system would give
exactly the same distribution as considered in this paper.

Linked to this, a further possibility that we are presently
investigating would be to integrate over the CM emission angle instead
of the Lab one considered in this work. This procedure would have the
advantage of being free of the divergence coming from the Jacobian,
but has the drawback of requiring to perform a boost back to the Lab
system for each pair of holes ($\nh_1,\nh_2$). One should perform a
full calculation with both approaches to see the advantages of each
one in terms of CPU time.

One of the main problems associated with a complete, exact calculation
of the 2p-2h response functions is the computational time required
when the full current is included. One of the outcomes of this work is
the possibility opened by considering what we called the frozen
nucleon approximation to compute the integral over the two holes.  The
validity of this approximation must be verified in the complete
calculation. If the approximation is found to be accurate enough, then
the calculation of the 2p-2h cross section could be done without much
difficulty and could be easily implemented in Monte-Carlo generators.
The verification of this approximation is one of the goals of our
future work.  Preliminary results obtained with the seagull diagrams
show that the approximation is valid for this set of diagrams.

The MC generators must not perform the integration over the outgoing
final state but instead must keep these momenta explicitly because one
is interested in generating a full final state to be propagated.  The
integration performed here is only needed for the inclusive 2p-2h cross
section, that cannot be separated from the measured QE cross section
if the final nucleons are not detected. With our model there is the
possibility to generate angular distributions of nucleon 2p-2h states
 produced by MEC, fully compatible with the inclusive 2p-2h 
cross sections. They could be useful for the MC generators.

\section{Conclusions}

We have performed a detailed study of the two-particle two-hole
phase-space function, which is proportional to the nuclear
two-particle emission response function for constant current matrix
elements. In order to obtain physically meaningful results one should
include a model for the two-body currents inside the integral.
However, the knowledge gained here by disregarding the operator and
focusing on the purely kinematical properties has been of great help
in optimizing the computation of the 7D integral appearing in the
2p-2h response functions of electron and neutrino scattering.  The
frozen nucleon approximation, that is, neglecting the momenta of the
initial nucleons for high momentum transfer, has allowed us to focus
on the angular distribution function. We have found that this function
has divergences for some angles. Our main goal has been to find the
allowed angular regions and to integrate analytically around the
divergent points. The CPU time of the 7D integral has been reduced
significantly.  The relativistic results converge to the
non-relativistic ones for low energy transfer. 
We are presently working on an implementation of the present
method with a complete model of the MEC operators.

\section*{Acknowledgments}
This work was supported by DGI (Spain): FIS2011-24149 and FIS2011-28738-C02-01, by the
Junta de Andaluc\'{\i}a (FQM-225 and FQM-160), by the Spanish Consolider-Ingenio 2010
programmed CPAN, in part (MBB) by the INFN project MANYBODY, and in part (TWD) by U.S.  Department of Energy under cooperative
agreement DE-FC02-94ER40818.  C.A. is supported by a CPAN postdoctoral
contract.


\appendix

\section{Calculation of $x_{max}$}

Here we derive the upper limit of the integral over $x$ in
Eq.~(\ref{fasico2D}).  We first note that the function $A(x,y,\nu)$
inside the integral contains the energy delta function 
$\delta(\omega+E_1+E_2-E'_1-E'_2)$ and the step function 
$\theta(k_F-h_1)\theta(k_F-h_2)$. 
This implies that
\begin{equation}
E'_1 
\leq  E'_1+E'_2 
= \omega+E_1+E_2 \leq \omega+2E_F\,.
\end{equation}
Therefore
\begin{equation}
\frac{p'_1{}^2}{2m_N}
\leq
\omega + 2\frac{k_F^2}{2m_N}\,.
\end{equation}
Taking the square root and rearranging, one has
\begin{equation}
p'_1 \leq k_F\sqrt{2+2m_N\omega/k_F^2}\,.
\end{equation}
Recalling now the definition of the non-dimensional variable
$\nu=m_N\omega/k_F^2$, we have
\begin{equation}
\frac{p'_1}{k_F}\leq\sqrt{2(1+\nu)}\,.
\end{equation}
Finally, using this inequality in the definition of the $x$ variable,
one finds that
\begin{equation}
x=\left|\frac{\np'_1-\nh_1}{k_F}\right| 
\leq
\frac{p'_1+h_1}{k_F}
\leq
\frac{p'_1+k_F}{k_F}
\leq
1+\sqrt{2(1+\nu)} \,.
\end{equation}

\section{The function $A(l_1,l_2,\nu)$}

The function $A(l_1,l_2,\nu)$ was computed analytically in
\cite{Van80}. In this work we have repeated the analytical
calculation and we have found a typographical error (a minus sign) in that
reference. Although the demonstration and numerical results of
\cite{Van80} are correct, taking into account the given error is essential.
For completeness, and because the error can mislead the reader,
we write in this appendix the correct final 
expression with the slightly different notation used by us.
We write the function $A$ as the sum of sixteen terms
\begin{equation}
A(l_1,l_2,\nu)
=
\sum_{i=1}^4
\sum_{j=1}^4
A_{ij}(l_1,l_2,\nu) \,,
\end{equation}
where the  $A_{ij}$ functions have the symmetry
\begin{equation}
A_{ij}(l_1,l_2,\nu)=
A_{ji}(l_2,l_1,\nu) \,.
\end{equation}
Thus, we only need to give the analytical expressions for the diagonal
and the upper half off-diagonal $ij$ elements
\begin{eqnarray}
A_{11}
&=& 
\left[ 
l_1l_2\frac{C_{11}^3}{3!}
+(l_1+l_2)\frac{C_{11}^4}{4!}
+\frac{C_{11}^5}{5!}
\right]
\theta(C_{11})
\nonumber\\
C_{11}
&\equiv&
\nu-\frac{l_1^2}{2}-l_1-\frac{l_2^2}{2}-l_2
\nonumber\\
A_{12}
&=& 
\left[ 
l_1l_2\frac{C_{12}^3}{3!}
+(l_2-l_1)\frac{C_{12}^4}{4!}
-\frac{C_{12}^5}{5!}
\right]
\theta(C_{12})\theta(l_2-2)
\nonumber\\
C_{12}
&\equiv&
\nu-\frac{l_1^2}{2}-l_1-\frac{l_2^2}{2}+l_2
\nonumber\\
A_{13}
&=& 
-\left[ 
l_1l_2\frac{C_{13}^3}{3!}
+(l_1+l_2)\frac{C_{13}^4}{4!}
+\frac{C_{13}^5}{5!}
\right]
\theta(C_{13})\theta(2-l_2)
\nonumber\\
C_{13}
&\equiv&
\nu-\frac{l_1^2}{2}-l_1+\frac{l_2^2}{2}-l_2
\nonumber\\
A_{14}
&=& 
\left[ 
l_1\frac{C_{14}^3}{3!}
+\frac{C_{14}^4}{4!}
\right]
l_2^2 \theta(C_{14})\theta(2-l_2)
\nonumber\\
C_{14}
&\equiv&
\nu-\frac{l_1^2}{2}-l_1
\nonumber\\
A_{22}
&=& 
\left[ 
l_1l_2\frac{C_{22}^3}{3!}
-(l_1+l_2)\frac{C_{22}^4}{4!}
+\frac{C_{22}^5}{5!}
\right]
\nonumber\\
&&
\times
\theta(C_{22})
\theta(l_1-2)
\theta(l_2-2)
\nonumber\\
C_{22}
&\equiv&
\nu-\frac{l_1^2}{2}+l_1-\frac{l_2^2}{2}+l_2
\nonumber\\
A_{23}
&=& 
-\left[ 
l_1l_2\frac{C_{23}^3}{3!}
+(l_1-l_2)\frac{C_{23}^4}{4!}
-\frac{C_{23}^5}{5!}
\right]
\nonumber\\
&&
\times
\theta(C_{23})
\theta(l_1-2)
\theta(2-l_2)
\nonumber\\
C_{23}
&\equiv&
\nu-\frac{l_1^2}{2}+l_1+\frac{l_2^2}{2}-l_2
\nonumber\\
A_{24}
&=& 
\left[ 
l_1\frac{C_{24}^3}{3!}
-\frac{C_{24}^4}{4!}
\right]
l_2^2
\theta(C_{24})
\theta(l_1-2)
\theta(2-l_2)
\nonumber\\
C_{24}
&\equiv&
\nu-\frac{l_1^2}{2}+l_1
\nonumber\\
A_{33}
&=& 
\left[ 
l_1l_2\frac{C_{33}^3}{3!}
+(l_1+l_2)\frac{C_{33}^4}{4!}
+\frac{C_{33}^5}{5!}
\right]
\nonumber\\
&&
\times
\theta(C_{33})
\theta(2-l_1)
\theta(2-l_2)
\nonumber\\
C_{33}
&\equiv&
\nu+\frac{l_1^2}{2}-l_1+\frac{l_2^2}{2}-l_2
\nonumber\\
A_{34}
&=& 
-\left[ 
l_1\frac{C_{34}^3}{3!}
+\frac{C_{34}^4}{4!}
\right]
l_2^2
\theta(C_{34})
\theta(2-l_1)
\theta(2-l_2)
\nonumber\\
C_{34}
&\equiv&
\nu+\frac{l_1^2}{2}-l_1
\nonumber\\
A_{44}
&=& 
l_1^2l_2^2\frac{\nu^3}{3!}
\theta(\nu)
\theta(2-l_1)
\theta(2-l_2)
\end{eqnarray}

Note that the equivalent function $A_{13}$ in \cite{Van80} 
(denoted $F_3$) has a missing global minus sign.

\section{Solutions of relativistic energy conservation. }

Given $q,\omega$ and
fixing the momenta of the two holes $\nh_1$, $\nh_2$, the total 
energy and momentum of the two particles is also fixed by
\begin{eqnarray*}
E' &=& E_1+E_2+\omega \\
\np' & = & \nh_1+\nh_2+\nq \,.
\end{eqnarray*}
For fixed emission angles of the first particle,
$\phi'_1=0$ and 
$\theta'_1$, 
the value of $p'_1$ is restricted 
by momentum conservation 
$\np'_2= \np'-\np'_1$
and energy conservation, $E'_2=E'-E'_1$.
In fact, taking the square of the last equation, we should solve
\begin{equation}
E'_2{}^2= (E'-E'_1)^2 \,.
\end{equation}
Having squared, we have introduced spurious solutions with $E'-E'_1<0$,
that should be thrown away.
Expanding the right-hand side, using the energy-momentum relation in 
the squared energies, and rearranging terms 
we arrive to the equivalent equation
\begin{equation} \label{energia-alternativa}
E'_1= \ta + \tv p'_1 \,,
\end{equation}
where we have defined
\begin{eqnarray}
\ta &=& \frac{E'{}^2-p'{}^2}{2E'} \label{ta} \\
\tv &=& \frac{\np'\cdot\hp'_1}{E'} \,. \label{tv}
\end{eqnarray}
Taking the square of Eq.~(\ref{energia-alternativa}) and again using
the energy-momentum relation we arrive at the second-degree equation
for $p'_1$
\begin{equation}
\tb p'_1{}^2 -2\ta\tv p'_1 +(m_N^2-\ta^2) = 0 \,,
\end{equation}
where we have defined
\begin{equation} \label{tb}
\tb= 1 - \tv^2.   
\end{equation}
Note that Eq.~(\ref{energia-alternativa}) provides an alternative way
to compute the energy $E'_1$ once $p'_1$ is known. It also is valid as
a check of the solution. However caution is needed because taking the
square of Eq.~(\ref{energia-alternativa}) introduces spurious
solutions with $ \ta + \tv p'_1 < 0 $ that should be disregarded.


\end{document}